\documentclass{aastex701}
\usepackage{graphicx} 
\usepackage{amsmath}
\received{May 24, 2026}
\accepted{September 18, 2026}
\submitjournal{PASP}

\makeatletter
\let\frontmatter@title@above=\relax
\makeatother

\begin{document}

\title{Phantom Zernike Wavefront Sensor: Phase Contrast with Engineered Focal Plane Reference Speckles}

\correspondingauthor{Benjamin Courtney-Barrer}
\author[0000-0002-7679-9686]{Benjamin Courtney-Barrer}
\affiliation{Australian Astronomical Optics, Macquarie University, NSW, Australia}
\affiliation{The Astrophysics and Space Technologies Research Centre, School of Mathematical and Physical Sciences, Macquarie University, NSW, Australia}
\affiliation{Research School of Astronomy and Astrophysics, Australian National University, Canberra 2611, Australia}
\email[show]{benjamin.courtney-barrer@mq.edu.au}




\begin{abstract}
We introduce the Phantom Zernike Wavefront Sensor (phantom ZWFS), an engineered reference extension of phase-contrast wavefront sensing in which controlled pupil plane modulation generates off-axis coherent speckle structures that undergo localized focal plane phase shifts. These references may replace the central ZWFS reference when the on-axis PSF core is required for another optical function, such as photonic science coupling, or may augment a conventional central ZWFS response to provide additional wavefront information. Using a monochromatic scalar model, we derive the resulting carrier structure and show how the engineered references can be recovered through spatial demodulation and incorporated into calibrated reconstruction. A simple four-reference realization demonstrates that the reference geometry redistributes photon sensitivity across spatial frequency magnitude and orientation, producing mode dependent changes in linearity and cross-talk rather than a uniform improvement over a classical ZWFS. We then consider differential piston on a seven segment aperture as an application specific example. A hybrid sensor retaining the central ZWFS phase region and adding six engineered references that match the segment geometry gives similar precision near the phased state but substantially improved performance at larger piston excursions. Under the adopted photon-noise-limited benchmark, the hybrid reduces the nonlinear reconstruction  error and provides better conditioned optical encoding of the piston state than an idealized vector ZWFS, allowing more precise piston estimation from the same detected photon flux at larger piston excursions.
\end{abstract}


\section{Introduction}
High performance wavefront sensing remains a central requirement for adaptive optics, high contrast imaging, and precision astronomical instrumentation. Among existing architectures, the Zernike wavefront sensor (ZWFS) \citep{zernike_1934_zwfs} is especially attractive because it converts small phase errors into an intensity signal through a focal plane phase shift of the focused light, and is widely recognized for its optimal photon noise sensitivity in the small aberration regime \citep{Guyon2005}. This optimal photon sensitivity has motivated its adoption and continued development for demanding astronomical applications \citep{bloemhof_2003_zwfs, ndiaye_2013_zelda,ndiaye_2018_III_zelda_onsky,haffert_2022_gmagaoX, courtney-barrer_2024_baldr,salama_2024_vzwfs}. At the same time, the classical ZWFS is not the endpoint of phase contrast sensing. More general Fourier based wavefront sensing frameworks have shown that the ZWFS can be understood as one member of a broader class of sensors that convert phase into measurable intensity structure through known optical filtering or modulation \citep{Fauvarque_2016_generalize_fourier_wfs}. Within this broader context, one route has been to retain the basic on-axis phase contrast geometry of the ZWFS while modifying its design parameters, diversity, or reconstruction strategy. Examples include modal optimization of design parameters for the ZWFS \citep{chambouleyron_2021_variationZWFS}, wavelength diversity \citep{darcis_2025_colorZWFS}, vector-Zernike implementations, which exploit polarization dependent phase shifts to enable simultaneous phase and amplitude sensing \citep{doelman_2019_vzwfs}, and phase-shifted Zernike variants, which generalize the focal plane phase bias in order to improve robustness and reconstruction performance outside the strict small signal limit \citep{wallace_2011_vzwfs,cisse_2022_phaseshifted_zwfs}.

A second route has been to introduce deliberately engineered reference structures rather than relying only on a single central phase dimple. In holographic and coronagraphic modal wavefront sensors, optically generated off-axis point spread function copies encode aberration information in dedicated focal plane channels \citep{wilby_2017_coronagraphic_modal_wavefront_sensor}. Another important example is the self-coherent camera (SCC), which uses the coherence of starlight to generate spatial fringes in the science focal plane through interference with a reference beam \citep{baudoz_2006_SCC_origin, galicher_2010_self-coherent_camera}. In that case, the fringes encode the complex speckle field and enable focal plane estimation of aberrations for wavefront control and high contrast imaging.  Temporal focal plane diversity provides a complementary approach, with pair-wise probing and the related Bord\'e--Traub method using known deformable mirror probe fields to recover the complex focal plane electric field \citep{Borde2006}.

A related motivation has emerged in high contrast coronagraphy and integrated optical systems, where wavefront sensing may need to share an optical train with the science measurement. For example, \citet{haffert_2023_PIAACMC} describe how the Phase Induced Amplitude Apodized Complex Mask Coronagraph (PIAACMC) can be combined with both a pupil plane Zernike wavefront sensor and a self coherent camera type focal plane estimator. \citet{taras_2026} recently provided extensions of this concept. Coronagraph based wavefront sensors provide another adjacent route, using the connection between high Strehl wavefront sensing and coronagraphic mode rejection to develop architectures that can combine science and sensing functions within a shared optical train \citep{chambouleyron2024_coronagraph_zwfs}. A further adjacent concept is the asymmetric pupil Fourier wavefront sensor (APF-WFS), in which wavefront errors are inferred from the Fourier analysis of a single direct image in the high Strehl regime, with pupil asymmetry introduced to make the inverse problem non-degenerate \citep{martinache_2013_apfwfs}.

Taken together, these developments point toward a broader design space in which phase contrast wavefront sensing can be constructed from deliberately generated and phase-biased coherent reference structures. Motivated by this perspective, we introduce here the Phantom Zernike Wavefront Sensor (phantom ZWFS), an engineered reference extension of Zernike-like phase contrast sensing. The defining feature of the architecture is the generation of controlled off-axis coherent speckle structures whose positions, strengths, and phase shifts can be adaptively changed, or selected as part of the sensor design. In the implementations considered here these references are generated by a weak pupil plane phase modulation, although the broader concept is not restricted to a sinusoidal waffle pattern. Localized focal plane phase shifts are applied to the engineered copies, and their interference with the remaining field is measured in a downstream pupil plane.

Importantly, the treatment of the physical on-axis PSF core is independent of the engineered reference channels. In one realization, the off-axis references provide the primary phase-contrast measurement while the central core is available for another optical function, such as coupling into a science waveguide or photonic chip. In another realization, the conventional central ZWFS phase shift is retained and the engineered references are added as auxiliary channels that provide additional optical diversity. The phantom ZWFS should therefore be viewed as a framework in which the coherent reference geometry itself becomes an additional degree of freedom in the design of the phase-contrast sensing operator.

This distinguishes the phantom approach from a classical or phase-shifted ZWFS, in which the sensing reference is defined primarily by the physical on-axis focal plane core. It also differs from SCC-like and holographic focal plane estimators in that the engineered reference copies undergo localized phase shifts and are read out through the resulting downstream pupil plane intensity. Their known spatial carrier frequencies can then be used for demodulation and reconstruction. The central question explored in this work is whether engineered coherent references can redistribute or augment the available wavefront information in ways that are useful for a particular sensing problem.

The remainder of the paper develops this idea in two stages. Section~\ref{sec:theory_reconstruction} introduces the generalized optical model, derives the weak modulation carrier description, and defines the direct and demodulated observables used for reconstruction. Section~\ref{sec:four_reference_characterization} then characterizes a simple standalone four-reference realization, including its photon-noise-limited Fourier sensitivity and the angular anisotropy imposed by the discrete reference geometry. Section~\ref{sec:benchmark_classical_vs_phantom} examines how this structured response appears in a calibrated Zernike-mode reconstruction relative to a classical ZWFS. Section~\ref{sec:segment_piston_example} then considers an application specific hybrid realization for differential piston on a seven-segment aperture, in which six engineered references augment rather than replace the central ZWFS response. Section~\ref{sec:discussion} discusses the resulting design trade space and limitations, and Section~\ref{sec:conclusion} summarizes the principal conclusions.

\section{Forward model, linearized response, and reconstruction}
\label{sec:theory_reconstruction}

\begin{figure}
    \centering
    \includegraphics[width=0.8\linewidth]{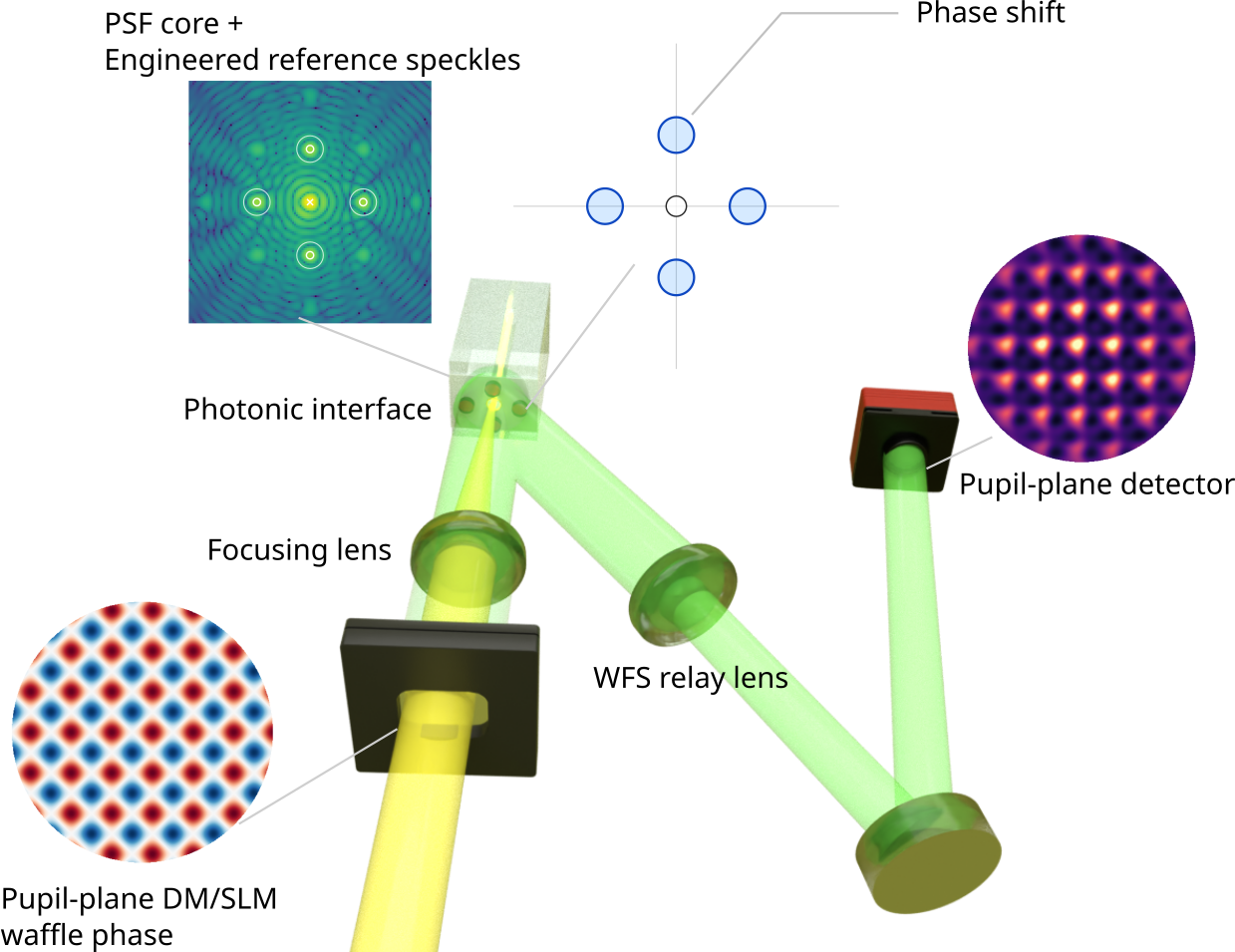}
    \caption{
    One core preserving realization of the phantom ZWFS concept. A pupil plane phase modulation, generated for example by a deformable mirror (DM), spatial light modulator (SLM), or fixed phase element, produces engineered first-order satellite speckles around the central PSF core. In the realization shown here, the core remains available for science coupling into a waveguide or photonic chip, while the satellite speckles encounter localized phase shifts and are returned to a downstream pupil plane detector. More generally, the central PSF core need not be removed from the sensing path: it may instead remain unmodified or undergo a conventional ZWFS phase shift, allowing the engineered satellite references to augment a central phase-contrast measurement. The returned pupil intensity contains spatial carriers at the known modulation frequencies, which can be demodulated to recover additional wavefront observables.
    }
    \label{fig:phantom_zwfs_concept_alg}
\end{figure}

The defining feature of the phantom ZWFS is the use of deliberately generated off-axis speckle structures as phase-shifted coherent reference channels. The treatment of the central PSF core is a separate design choice. In one realization, illustrated in Fig.~\ref{fig:phantom_zwfs_concept_alg}, the central core is coupled into a science waveguide or photonic system and only the non-coupled light is used for wavefront sensing. In another realization, developed later in this paper, the central core remains in the wavefront sensing path and undergoes a conventional ZWFS phase shift, while engineered off-axis references provide additional measurements. The phantom architecture can therefore operate either as a standalone phase-contrast sensor or as an auxiliary reference system that augments an existing central phase-contrast measurement.

We first formulate the optical propagation in a way that encompasses both cases. We then specialize to the simple four-reference waffle geometry used for the baseline characterization in Sections~3 and~4. For the forward model we work with a scalar monochromatic field sampled on a discrete pupil grid, with pupil plane coordinate $\mathbf{r}=(x,y)$ and focal plane coordinate $\mathbf{u}$. The derivation is organized in three layers: an exact field-propagation model, a weak waffle carrier expansion used to expose the signal structure, and a first-order linearization in the unknown aberration. The exact numerical simulations retain the full exponential pupil modulation; the weak waffle expansion is used only to interpret the resulting carrier structure.

\subsection{General forward model}
\label{sec:forward_model}

We begin with the incident pupil field
\begin{equation}
E_{\mathrm{in}}(\mathbf{r})
=
P(\mathbf{r})
\exp\!\left[i\phi(\mathbf{r})\right],
\end{equation}
where $P(\mathbf{r})$ is the pupil transmission and $\phi(\mathbf{r})$ is the unknown phase aberration to be sensed. An engineered pupil plane phase modulation $\phi_{\mathrm{w}}(\mathbf{r})$ is then applied. More generally this may contain several carrier directions,
\begin{equation}
\phi_{\mathrm{w}}(\mathbf{r})
=
\sum_{\ell=1}^{N_{\mathrm{c}}}
a_{\ell}
\sin\!\left(
2\pi \mathbf{f}_{\ell}\cdot\mathbf{r}
+
\psi_{\ell}
\right),
\label{eq:general_waffle}
\end{equation}
where $\mathbf{f}_{\ell}$, $a_{\ell}$, and $\psi_{\ell}$ specify the spatial frequency, amplitude, and phase of each pupil carrier. Such a modulation may be introduced using a fixed optical element, deformable mirror, or spatial light modulator. The field entering the sensing architecture is therefore
\begin{equation}
E_{0}(\mathbf{r})
=
P(\mathbf{r})
\exp\!\left[
i\bigl(
\phi(\mathbf{r})
+
\phi_{\mathrm{w}}(\mathbf{r})
\bigr)
\right].
\label{eq:E0_exact}
\end{equation}
Its focal plane field is
\begin{equation}
\widetilde{E}_{0}(\mathbf{u})
=
\mathcal{F}
\!\left\{
E_{0}(\mathbf{r})
\right\},
\end{equation}
where $\mathcal{F}$ denotes a centered Fourier transform. We separate two operations that may occur in this focal plane. First, let $\mathcal{R}$ denote a linear routing or filtering operator that determines which part of the focal plane field reaches the wavefront sensing branch,
\begin{equation}
\widetilde{E}_{\mathrm{wfs}}(\mathbf{u})
=
\mathcal{R}
\!\left[
\widetilde{E}_{0}
\right](\mathbf{u}).
\label{eq:general_routing}
\end{equation}
For a sensor in which all light remains in the wavefront sensing path, $\mathcal{R}$ is simply the identity operator. Alternatively, $\mathcal{R}$ may describe partial coupling of the central core into a science channel, reflection from a photonic interface, or another fixed linear optical filtering operation.

Second, the focal plane field may encounter both a central phase-shifting region and one or more localized phase shifts centred on the engineered satellite references. We write
\begin{equation}
M_{0}(\mathbf{u})
=
\exp\!\left[
i\Phi_{0}m_{0}(\mathbf{u})
\right]
\end{equation}
for an optional central phase mask, and
\begin{equation}
M_{\mathrm{s}}(\mathbf{u})
=
\prod_{j=1}^{N_{\mathrm{s}}}
\exp\!\left[
i\Phi_j
m_j(\mathbf{u}-\mathbf{u}_j)
\right]
\label{eq:Ms_general}
\end{equation}
for the satellite phase-shifting regions. Here $m_0$ and $m_j$ describe the spatial supports of the central and satellite phase regions and $\Phi_0$ and $\Phi_j$ their respective phase shifts. The downstream pupil plane field is then
\begin{equation}
E_{\mathrm{out}}(\mathbf{r})
=
\mathcal{F}^{-1}
\!\left\{
M_{0}(\mathbf{u})
M_{\mathrm{s}}(\mathbf{u})
\,
\mathcal{R}
\!\left[
\widetilde{E}_{0}
\right](\mathbf{u})
\right\}.
\label{eq:Eout_general}
\end{equation}
The measured intensity is
\begin{equation}
I_{\mathrm{out}}(\mathbf{r};\phi)
=
\left|
E_{\mathrm{out}}(\mathbf{r})
\right|^2.
\end{equation}
This notation contains several useful limiting cases. A classical ZWFS is recovered by setting $\phi_{\mathrm{w}}=0$, $M_{\mathrm{s}}=1$, and using a central phase region $M_0$ with the usual ZWFS phase shift. A standalone phantom ZWFS instead uses non-zero $\phi_{\mathrm{w}}$ and $M_{\mathrm{s}}$, while $M_0=1$; the central PSF core may then be removed from the sensing path through $\mathcal{R}$ or simply left unmodified. A hybrid configuration retains both $M_0$ and the engineered satellite references, allowing the conventional ZWFS signal and the phantom carrier channels to be measured simultaneously.

To isolate the aberration dependent signal, we define a reference measurement with the same pupil modulation, routing operator, and focal plane phase masks, but zero input aberration,
\begin{equation}
I_{\mathrm{ref}}(\mathbf{r})
=
I_{\mathrm{out}}(\mathbf{r};\phi=0).
\end{equation}
The differential pupil intensity is therefore
\begin{equation}
\Delta I(\mathbf{r})
=
I_{\mathrm{out}}(\mathbf{r};\phi)
-
I_{\mathrm{ref}}(\mathbf{r}).
\label{eq:deltaI_def}
\end{equation}
Equations~(\ref{eq:E0_exact})--(\ref{eq:deltaI_def}) define the general monochromatic forward model.

\subsection{Core preserving photonic realization}
\label{sec:photonic_realization}

The architecture illustrated in Fig.~\ref{fig:phantom_zwfs_concept_alg} corresponds to the particular case in which the central PSF core is used as a science input to a single mode waveguide or photonic chip. In this realization the central phase mask is absent,
$M_0=1$, and the wavefront sensing reference is provided by the engineered satellite speckles. The routing operator $\mathcal{R}$ then describes partial removal of the field through coupling into the guided mode.

Let $g(\mathbf{u})$ denote the normalized guided mode of the waveguide,
\begin{equation}
\int
|g(\mathbf{u})|^2
\,d\mathbf{u}
=
1.
\end{equation}
For an arbitrary focal plane field $\widetilde{E}$, the coupled complex amplitude is
\begin{equation}
a_{\mathrm{c}}
=
\int
\widetilde{E}(\mathbf{u})
g^{*}(\mathbf{u})
\,d\mathbf{u},
\end{equation}
and the corresponding coupled field component is
\begin{equation}
\widetilde{E}_{\mathrm{c}}(\mathbf{u})
=
a_{\mathrm{c}}g(\mathbf{u}).
\end{equation}

Let $c(\mathbf{u})$ denote the spatial support of the chip input region. The routing operator adopted for the photonic realization is
\begin{equation}
\mathcal{R}_{\mathrm{chip}}
\!\left[
\widetilde{E}
\right](\mathbf{u})
=
r\left[
(1-c(\mathbf{u}))
\widetilde{E}(\mathbf{u})
+
c(\mathbf{u})
\left(
\widetilde{E}(\mathbf{u})
-
\widetilde{E}_{\mathrm{c}}(\mathbf{u})
\right)
\right],
\label{eq:Erefl_chip}
\end{equation}
where $r$ is the normal-incidence Fresnel amplitude reflection coefficient at the chip input interface. In this simplified model, all light outside the coupling region remains available to the sensing branch, while within the coupling region only the component that fails to couple to the guided mode is returned. This overlap integral treatment is intended to capture the principal photon-sharing geometry and is not a full vector scattering model of the photonic interface.

The resulting architecture is useful when the diffraction-limited central core is itself the desired science signal. The off-axis references provide a separate phase-contrast sensing resource whose number, location, strength, and phase shifts may be chosen independently of the science coupling. Related approaches in high-contrast instrumentation also combine science and wavefront sensing functions within a shared optical train \citep{chambouleyron2024_coronagraph_zwfs}. The distinction here is that the phantom architecture deliberately generates off-axis coherent speckle structures and applies localized phase shifts to those references rather than relying only on manipulation of the physical on-axis core.

The photonic realization is the configuration used for the standalone phantom characterization in Sections~3 and~4. The later hybrid segment-piston example instead uses $\mathcal{R}=1$ and retains a conventional central ZWFS phase region together with the engineered satellite references.

\subsection{Weak waffle carrier expansion}
\label{sec:weak_waffle}
The general forward model above permits arbitrary carrier geometries. To expose the carrier structure analytically, we now specialize to the simple four-reference modulation used for the baseline phantom ZWFS characterization,
\begin{equation}
\phi_{\mathrm{w}}(\mathbf{r})
=
a
\left[
\sin(2\pi f x)
+
\cos(2\pi f y)
\right].
\label{eq:four_spot_waffle}
\end{equation}
This produces first-order satellite references along the $\pm x$ and $\pm y$ directions. The later hybrid application uses the same principle with three carrier axes rather than two.

Define the waffle-biased reference field
\begin{equation}
E_{\mathrm{w}}(\mathbf{r})
=
P(\mathbf{r})
e^{i\phi_{\mathrm{w}}(\mathbf{r})}.
\end{equation}
For weak waffle amplitude, $a\ll1$, the Jacobi--Anger expansions
\begin{equation}
e^{ia\sin X}
=
\sum_{n=-\infty}^{\infty}
J_n(a)e^{inX},
\qquad
e^{ia\cos Y}
=
\sum_{m=-\infty}^{\infty}
i^mJ_m(a)e^{imY},
\end{equation}
give
\begin{equation}
e^{i\phi_{\mathrm{w}}(\mathbf{r})}
=
\sum_{n,m}
J_n(a)i^mJ_m(a)
e^{i2\pi(nfx+mfy)}.
\end{equation}

Retaining only terms through first order in $a$ gives
\begin{equation}
e^{i\phi_{\mathrm{w}}(\mathbf{r})}
\approx
1
+
\frac{a}{2}e^{i2\pi fx}
-
\frac{a}{2}e^{-i2\pi fx}
+
\frac{ia}{2}e^{i2\pi fy}
+
\frac{ia}{2}e^{-i2\pi fy}.
\label{eq:weak_waffle_expansion}
\end{equation}
Thus the waffle-biased field contains a zero-order component together with four retained first-order carrier bands at spatial frequencies $(\pm f,0)$ and $(0,\pm f)$,
\begin{equation}
E_{\mathrm{w}}(\mathbf{r})
\approx
\sum_{q\in\mathcal Q}
\beta_q
P(\mathbf{r})
e^{i2\pi\mathbf{k}_q\cdot\mathbf{r}},
\label{eq:Ew_band_sum}
\end{equation}
where
\begin{equation}
\mathcal Q
=
\{0,x+,x-,y+,y-\},
\end{equation}
with carrier vectors
\begin{align}
\mathbf{k}_{0}&=(0,0), &
\mathbf{k}_{x+}&=(f,0), &
\mathbf{k}_{x-}&=(-f,0), &
\mathbf{k}_{y+}&=(0,f), &
\mathbf{k}_{y-}&=(0,-f),
\end{align}
and coefficients
\begin{align}
\beta_{0}&=1, &
\beta_{x+}&=\frac{a}{2}, &
\beta_{x-}&=-\frac{a}{2}, &
\beta_{y+}&=\frac{ia}{2}, &
\beta_{y-}&=\frac{ia}{2}.
\end{align}

Physically, the pupil modulation decomposes the field into a known carrier set whose shifted focal plane copies form the engineered reference speckles. Higher-order harmonics are neglected only in this analytic interpretation and are discussed further in Appendix~\ref{sec:appendix_waffle}. All numerical simulations in this paper propagate the full exponential modulation of Eq.~(\ref{eq:E0_exact}).

\subsection{Linearization in the unknown aberration}
\label{sec:linearization_aberration}
We next linearize in the unknown phase aberration while retaining the fixed engineered reference geometry. Writing
\begin{equation}
E_0(\mathbf{r})
=
E_{\mathrm{w}}(\mathbf{r})
e^{i\phi(\mathbf{r})},
\end{equation}
and assuming $|\phi(\mathbf{r})|\ll1$, we use
\begin{equation}
e^{i\phi(\mathbf{r})}
\approx
1+i\phi(\mathbf{r})
\end{equation}
to obtain
\begin{equation}
E_0(\mathbf{r})
\approx
E_{\mathrm{w}}(\mathbf{r})
+
iE_{\mathrm{w}}(\mathbf{r})
\phi(\mathbf{r}).
\label{eq:E0_linearized}
\end{equation}
Substituting the retained carrier expansion of Eq.~(\ref{eq:Ew_band_sum}) gives
\begin{equation}
E_0(\mathbf{r})
\approx
\sum_{q\in\mathcal Q}
\beta_qP(\mathbf{r})
e^{i2\pi\mathbf{k}_q\cdot\mathbf{r}}
+
i
\sum_{q\in\mathcal Q}
\beta_qP(\mathbf{r})
\phi(\mathbf{r})
e^{i2\pi\mathbf{k}_q\cdot\mathbf{r}}.
\label{eq:E0_band_sum}
\end{equation}
Let $\mathcal{L}$ denote the complete linear optical propagation from the pupil field to the returned pupil field,
\begin{equation}
\mathcal{L}[E]
=
\mathcal{F}^{-1}
\!\left\{
M_0
M_{\mathrm{s}}
\,
\mathcal{R}
\!\left[
\mathcal{F}\{E\}
\right]
\right\},
\label{eq:L_general}
\end{equation}
so that
\begin{equation}
E_{\mathrm{out}}(\mathbf{r})
=
\mathcal{L}[E_0](\mathbf{r}).
\end{equation}
The operator $\mathcal{L}$ therefore includes any central phase-contrast region, the engineered satellite phase shifts, focal plane routing or coupling, and propagation back to the pupil. Its linearity refers to its action on the complex optical field; the subsequent intensity measurement remains nonlinear.

In the weak waffle regime, and provided the retained carrier bands remain sufficiently separated after propagation, we approximate the contribution from each retained carrier as the same rapidly varying carrier multiplied by a slowly varying complex envelope,
\begin{align}
\mathcal{L}
\!\left[
\beta_qP
e^{i2\pi\mathbf{k}_q\cdot\mathbf{r}}
\right]
&\approx
B_q(\mathbf{r})
e^{i2\pi\mathbf{k}_q\cdot\mathbf{r}},
\label{eq:Bq_def}
\\
\mathcal{L}
\!\left[
i\phi\,
\beta_qP
e^{i2\pi\mathbf{k}_q\cdot\mathbf{r}}
\right]
&\approx
D_q(\mathbf{r})
e^{i2\pi\mathbf{k}_q\cdot\mathbf{r}}.
\label{eq:Dq_def}
\end{align}
Here $B_q$ are the zero aberration output envelopes associated with each retained carrier band and $D_q$ are the corresponding first-order perturbation envelopes. Because $\mathcal{L}$ is linear in the field and the second expression is first order in $\phi$, each $D_q$ is linear in the unknown aberration. The assumptions and limitations of this reduced carrier-envelope representation are discussed in Appendix~\ref{app:reduced_envelope}. The output field can therefore be written
\begin{equation}
E_{\mathrm{out}}(\mathbf{r})
\approx
E_{\mathrm{out}}^{(0)}(\mathbf{r})
+
\delta E_{\mathrm{out}}(\mathbf{r}),
\end{equation}
with
\begin{align}
E_{\mathrm{out}}^{(0)}(\mathbf{r})
&\approx
\sum_{q\in\mathcal Q}
B_q(\mathbf{r})
e^{i2\pi\mathbf{k}_q\cdot\mathbf{r}},
\label{eq:Eout0_band_sum}
\\
\delta E_{\mathrm{out}}(\mathbf{r})
&\approx
\sum_{q\in\mathcal Q}
D_q(\mathbf{r})
e^{i2\pi\mathbf{k}_q\cdot\mathbf{r}}.
\label{eq:dEout_band_sum}
\end{align}
The corresponding intensity is
\begin{equation}
I_{\mathrm{out}}(\mathbf{r};\phi)
\approx
\left|
E_{\mathrm{out}}^{(0)}(\mathbf{r})
\right|^2
+
2\,
\Re
\!\left[
\left(
E_{\mathrm{out}}^{(0)}(\mathbf{r})
\right)^*
\delta E_{\mathrm{out}}(\mathbf{r})
\right],
\end{equation}
and subtraction of the zero aberration reference gives
\begin{equation}
\Delta I(\mathbf{r})
\approx
2\,
\Re
\!\left[
\left(
E_{\mathrm{out}}^{(0)}(\mathbf{r})
\right)^*
\delta E_{\mathrm{out}}(\mathbf{r})
\right].
\label{eq:linear_signal}
\end{equation}
Substituting Eqs.~(\ref{eq:Eout0_band_sum}) and~(\ref{eq:dEout_band_sum}) yields
\begin{equation}
\Delta I(\mathbf{r})
\approx
2\,
\Re
\!\left[
\sum_{p,q\in\mathcal Q}
B_q^*(\mathbf{r})
D_p(\mathbf{r})
e^{i2\pi(\mathbf{k}_p-\mathbf{k}_q)\cdot\mathbf{r}}
\right].
\label{eq:deltaI_band_sum}
\end{equation}
Equation~(\ref{eq:deltaI_band_sum}) is the central result of the linearized carrier model. It shows that the first-order differential pupil signal consists of known spatial carrier terms whose slowly varying coefficients remain linear in the unknown aberration through the perturbation envelopes $D_q$. The engineered satellite references therefore modify the optical interference structure before detection rather than merely defining a post-processing basis.

\subsection{Carrier decomposition of the linearized signal}
Equation~(\ref{eq:deltaI_band_sum}) contains a zero-order term ($\mathbf{k}_p-\mathbf{k}_q=0$), first-order terms at $(\pm f,0)$ and $(0,\pm f)$, second-order terms at $(\pm2f,0)$ and $(0,\pm2f)$, and oblique cross terms such as $(\pm f,\mp f)$. The carrier reconstruction considered for the baseline four-reference phantom sensor retains the first-order components at $(\pm f,0)$ and $(0,\pm f)$. For the $+x$ carrier, the condition $\mathbf{k}_p-\mathbf{k}_q=(f,0)$ is satisfied by $(p,q)=(x+,0)$ and $(p,q)=(0,x-)$. Similarly, the $+y$ carrier contains the pairs $(p,q)=(y+,0)$ and $(p,q)=(0,y-)$. The corresponding first-order envelopes are
\begin{align}
\Gamma_x(\mathbf{r})
&=
B_0^*(\mathbf{r})
D_{x+}(\mathbf{r})
+
B_{x-}^*(\mathbf{r})
D_0(\mathbf{r}),
\label{eq:Gamma_x_def}
\\
\Gamma_y(\mathbf{r})
&=
B_0^*(\mathbf{r})
D_{y+}(\mathbf{r})
+
B_{y-}^*(\mathbf{r})
D_0(\mathbf{r}).
\label{eq:Gamma_y_def}
\end{align}
By construction, $\Gamma_x$ and $\Gamma_y$ are slowly varying over the pupil and remain linear in $\phi$. The retained first-order contribution to the real differential intensity can therefore be written
\begin{equation}
\Delta I_{\pm f}(\mathbf{r})
\approx
\Gamma_x(\mathbf{r})
e^{i2\pi fx}
+
\Gamma_x^*(\mathbf{r})
e^{-i2\pi fx}
+
\Gamma_y(\mathbf{r})
e^{i2\pi fy}
+
\Gamma_y^*(\mathbf{r})
e^{-i2\pi fy},
\label{eq:deltaI_complex_carriers}
\end{equation}
or equivalently
\begin{equation}
\Delta I_{\pm f}(\mathbf{r})
\approx
A_x(\mathbf{r})\cos(2\pi fx)
+
B_x(\mathbf{r})\sin(2\pi fx)
+
A_y(\mathbf{r})\cos(2\pi fy)
+
B_y(\mathbf{r})\sin(2\pi fy),
\label{eq:carrier_form_real}
\end{equation}
where
\begin{align}
A_x(\mathbf{r})
&=
2\Re[\Gamma_x(\mathbf{r})],
&
B_x(\mathbf{r})
&=
-2\Im[\Gamma_x(\mathbf{r})],
\\
A_y(\mathbf{r})
&=
2\Re[\Gamma_y(\mathbf{r})],
&
B_y(\mathbf{r})
&=
-2\Im[\Gamma_y(\mathbf{r})].
\end{align}
Thus $A_x$, $B_x$, $A_y$, and $B_y$ are the in-phase and quadrature envelopes associated with the known pupil carriers. The conditions under which these components can be isolated are summarized in Appendix~\ref{sec:appendix_demodulation}.

\subsection{Carrier demodulation}
The slowly varying envelopes in Eq.~(\ref{eq:carrier_form_real}) are estimated numerically by complex spatial demodulation,
\begin{align}
M_x(\mathbf{r})
&=
\mathrm{LPF}
\!\left[
\Delta I(\mathbf{r})
e^{-i2\pi fx}
\right],
\label{eq:Mx_def}
\\
M_y(\mathbf{r})
&=
\mathrm{LPF}
\!\left[
\Delta I(\mathbf{r})
e^{-i2\pi fy}
\right],
\label{eq:My_def}
\end{align}
where $\mathrm{LPF}$ denotes a low-pass spatial filter. Under the carrier-separation assumptions,
\begin{equation}
M_x(\mathbf{r})
\approx
\Gamma_x(\mathbf{r}),
\qquad
M_y(\mathbf{r})
\approx
\Gamma_y(\mathbf{r}),
\end{equation}
while the translated zero-order, second-order, and oblique terms are rejected. For the standalone four-reference phantom sensor studied in Sections~3 and~4, the reconstruction uses the carrier feature vector
\begin{equation}
\mathbf{y}_{\mathrm{sat}}
=
\begin{bmatrix}
\Re(M_x)\\
\Im(M_x)\\
\Re(M_y)\\
\Im(M_y)
\end{bmatrix}_{P>0},
\label{eq:y_sat}
\end{equation}
where the subscript denotes restriction to pixels within the illuminated pupil. The general architecture does not require the direct pupil plane signal to be discarded. If a central ZWFS phase region is retained, for example, the differential pupil intensity itself contains useful complementary information. We therefore also define
\begin{equation}
\mathbf{y}_{0}
=
\left[
\Delta I(\mathbf{r})
\right]_{P>0}.
\label{eq:y_direct}
\end{equation}
A hybrid measurement may then concatenate the direct and carrier observables,
\begin{equation}
\mathbf{y}_{\mathrm{hyb}}
=
\begin{bmatrix}
\mathbf{y}_{0}\\
\mathbf{y}_{\mathrm{sat}}
\end{bmatrix}.
\label{eq:y_hybrid_general}
\end{equation}
The carrier observables in Eq.~(\ref{eq:y_sat}) are deterministic transformations of the same detector image appearing in Eq.~(\ref{eq:y_direct}); they therefore do not constitute independent photon measurements. When a noise-weighted estimator is used, the corresponding correlations must be retained in the measurement covariance. This direct-plus-carrier construction is used later for the hybrid segment-piston example.

For a more general engineered modulation containing $N_{\mathrm{c}}$ carrier directions, the same procedure gives one complex demodulated observable $M_j$ for each retained carrier,
\begin{equation}
M_j(\mathbf{r})
=
\mathrm{LPF}
\!\left[
\Delta I(\mathbf{r})
e^{-i2\pi\mathbf{f}_j\cdot\mathbf{r}}
\right],
\end{equation}
so that the four-reference measurement above is simply the $N_{\mathrm{c}}=2$ case.

\subsection{Modal interaction matrix and reconstruction}
\label{sec:build_interaction_matrix}
The reconstruction procedure is independent of whether the chosen measurement vector contains only the demodulated satellite channels, the direct pupil intensity, or a combination of both. We denote the selected measurement vector generically by $\mathbf{y}$. The phase is represented in a modal basis,
\begin{equation}
\phi(\mathbf{r})
=
\sum_{k=1}^{N}
a_kZ_k(\mathbf{r}),
\end{equation}
where $Z_k$ are the reconstruction modes and $a_k$ their coefficients. In the baseline implementation the $Z_k$ are RMS-normalized Zernike modes on the sampled pupil. A linear interaction matrix $\mathbf{H}$ is constructed numerically using push--pull finite differences. For each calibration mode $Z_k$, positive and negative perturbations of amplitude $\epsilon$ are applied,
\begin{equation}
\phi_k^{\pm}(\mathbf{r})
=
\pm\epsilon Z_k(\mathbf{r}),
\end{equation}
the corresponding differential detector images are propagated through the selected measurement operator to produce $\mathbf{y}_k^{\pm}$, and the $k$th column of the interaction matrix is
\begin{equation}
\mathbf{H}_{:,k}
=
\frac{
\mathbf{y}_k^{+}
-
\mathbf{y}_k^{-}
}{
2\epsilon
}.
\label{eq:H_def}
\end{equation}
To first order,
\begin{equation}
\mathbf{y}
\approx
\mathbf{H}\mathbf{a}.
\end{equation}
For the baseline reconstruction, the estimated coefficients are obtained from the truncated Moore--Penrose pseudo-inverse,
\begin{equation}
\hat{\mathbf{a}}
=
\mathbf{H}^{+}\mathbf{y},
\end{equation}
and the reconstructed phase is
\begin{equation}
\hat{\phi}(\mathbf{r})
=
\sum_{k=1}^{N}
\hat{a}_kZ_k(\mathbf{r}).
\end{equation}
For a test phase represented in the same basis, the residual is
\begin{equation}
\phi_{\mathrm{res}}(\mathbf{r})
=
\phi(\mathbf{r})
-
\hat{\phi}(\mathbf{r}).
\end{equation}
The interaction-matrix formulation is deliberately general. The optical reference geometry determines how aberrations are encoded into the detector intensity, while the selected feature vector determines which components of that intensity are presented to the estimator. In the standalone four-reference phantom ZWFS, only the demodulated carrier observables of Eq.~(\ref{eq:y_sat}) are used. In a hybrid implementation, a conventional central phase-contrast response can instead be retained through Eq.~(\ref{eq:y_direct}) and augmented by additional engineered carrier channels.

In summary, the phantom ZWFS framework consists of: (i) generating controlled off-axis coherent reference channels through a known pupil modulation; (ii) applying localized phase shifts to those references, with the treatment of the central PSF core chosen independently for the application; (iii) measuring a differential downstream pupil intensity relative to a fixed zero aberration reference; and (iv) recovering wavefront information either from the direct phase-contrast signal, from spatially demodulated carrier channels, or from both simultaneously. The four-spot core preserving architecture characterized in the following sections is therefore one realization of a broader engineered reference phase-contrast framework.

\section{Characterization of a four-reference phantom ZWFS}
\label{sec:four_reference_characterization}
We now characterize one specific realization of the general engineered reference framework introduced in Section~\ref{sec:theory_reconstruction}: a standalone phantom ZWFS using two orthogonal pupil carriers to generate four phase-shifted satellite references along the $\pm x$ and $\pm y$ directions. In this configuration there is no central ZWFS phase region and we retain the core preserving photonic routing model of Section~\ref{sec:photonic_realization}, such that the central diffraction order may be coupled into a science channel while the remaining light is used for wavefront sensing.

The purpose of this section is not to identify a globally optimal phantom ZWFS geometry. The broader framework permits the number, orientation, strength, spatial frequency, and phase shift of the engineered references to be varied, and later in this paper we consider a different reference geometry for a segmented-pupil application. Instead, we use the simplest four-reference realization to establish two basic properties of engineered reference phase-contrast sensing: first, that the spatial distribution of the references redistributes sensitivity in Fourier space; and second, that the geometric parameters of the reference channels determine the photon efficiency of that particular realization.

\subsection{Photon-noise-limited sensitivity metric}
\label{sec:sensitivity_metric}
To quantify the small signal response of the four-reference phantom ZWFS and compare it with a conventional Zernike wavefront sensor, we adopt the photon-noise-limited Fourier mode sensitivity formalism of \citet{Guyon2005}. This metric is useful here because it characterizes the intrinsic response to a single spatial frequency perturbation, and because applying the same procedure to a classical ZWFS provides a direct reference against established wavefront sensing limits.

We consider a monochromatic pupil plane phase perturbation written as a Fourier cosine--sine pair,
\begin{equation}
\phi(\mathbf{r})
=
a_{\mathrm{c}}
\cos(2\pi\mathbf{f}\cdot\mathbf{r})
+
a_{\mathrm{s}}
\sin(2\pi\mathbf{f}\cdot\mathbf{r}),
\end{equation}
where $\mathbf{f}$ is the spatial frequency vector and
$a_{\mathrm{c}}$ and $a_{\mathrm{s}}$ are the cosine and sine coefficients.
In the small-aberration regime, the corresponding measurement vector is
\begin{equation}
\mathbf{y}
=
\mathbf{H}_{\mathbf{f}}
\begin{bmatrix}
a_{\mathrm{c}}\\
a_{\mathrm{s}}
\end{bmatrix}
+
\mathbf{n},
\end{equation}
where $\mathbf{H}_{\mathbf{f}}$ is the two-column block of the interaction matrix associated with the cosine and sine components of the Fourier mode at spatial frequency $\mathbf{f}$, constructed using the push--pull procedure of Section~\ref{sec:build_interaction_matrix}, and $\mathbf{n}$ denotes measurement noise. For a linear unbiased estimator, the covariance of the recovered Fourier coefficients is
\begin{equation}
\mathbf{C}_{\mathbf{f}}
=
\left(
\mathbf{H}_{\mathbf{f}}^{\mathsf T}
\mathbf{C}_{n}^{-1}
\mathbf{H}_{\mathbf{f}}
\right)^{-1},
\end{equation}
where $\mathbf{C}_{n}$ is the measurement noise covariance. If the diagonal terms of $\mathbf{C}_{\mathbf{f}}$ are
$\sigma_{\mathrm{c}}^2$ and $\sigma_{\mathrm{s}}^2$, we define the total Fourier mode estimation uncertainty as
\begin{equation}
\Sigma
=
\sqrt{
\sigma_{\mathrm{c}}^2
+
\sigma_{\mathrm{s}}^2
}.
\end{equation}
Following the normalization of \citet{Guyon2005}, the photon noise sensitivity coefficient $\beta_p$ is defined through
\begin{equation}
\beta_p
=
\Sigma\sqrt{N_{\mathrm{ph}}},
\end{equation}
where $N_{\mathrm{ph}}$ is the number of photons available to the wavefront sensing measurement. Smaller $\beta_p$ therefore corresponds to greater photon efficiency.

For the classical ZWFS reference, $N_{\mathrm{ph}}$ is the photon number incident on the sensing channel. For the core preserving phantom realization characterized in this section, only photons returned to the wavefront sensing branch are included,
\begin{equation}
N_{\mathrm{ph,WFS}}
=
\eta_{\mathrm{refl}}
N_{\mathrm{ph,inc}},
\end{equation}
where $\eta_{\mathrm{refl}}$ is the fraction of incident light remaining in the sensing path after science coupling. The resulting $\beta_p$ therefore measures how efficiently the photons delivered to the WFS encode the Fourier perturbation; the separate loss of incident flux through science coupling is accounted for by $\eta_{\mathrm{refl}}$. For context, the maximum theoretical coupling of an unaberrated Airy pattern into a single mode fibre is approximately $\eta_{\mathrm{coup}}\simeq0.78$ for an unobstructed circular aperture \citep{ShaklanRoddier1988}. For each spatial frequency and orientation, we construct the corresponding cosine and sine Fourier modes on the sampled pupil, obtain $\mathbf{H}_{\mathbf{f}}$ numerically from finite differences, and propagate the Poisson measurement covariance to $\mathbf{C}_{\mathbf{f}}$. The zero aberration reference image is treated as known exactly, so photon noise is applied only to the measured sensing image. This convention is used consistently for the classical and phantom calculations.

\subsection{Angular response of the four-reference geometry}
\label{sec:angular_anisotropy}

Before examining the geometric parameters of the four-reference realization, we first characterize its angular response in Fourier space. Unlike the rotationally symmetric classical ZWFS, the engineered reference field considered here is generated by two orthogonal pupil carriers, producing four discrete satellite references along the $x$ and $y$ axes. The resulting response is therefore expected to depend not only on the spatial frequency magnitude $|\mathbf{f}|$, but also on the orientation $\theta$ of the input Fourier mode relative to the reference geometry.

To quantify this dependence, we evaluate $\beta_p$ for Fourier phase modes at spatial frequencies
\begin{equation}
|\mathbf{f}|
=
\{2,4,6,8,10,12\}\,\lambda/D.
\end{equation}
For each spatial frequency, the orientation angle $\theta$, measured relative to the positive $x$ carrier axis, is sampled from $0^\circ$ to $90^\circ$ in $5^\circ$ increments, giving 19 orientations over the unique angular interval of the fourfold-symmetric geometry. For every $(|\mathbf{f}|,\theta)$ pair, the cosine - sine interaction matrix is constructed and its Poisson-limited coefficient covariance converted to $\beta_p$ using the procedure above.

\begin{figure}
    \centering
    \includegraphics[width=0.95\linewidth]{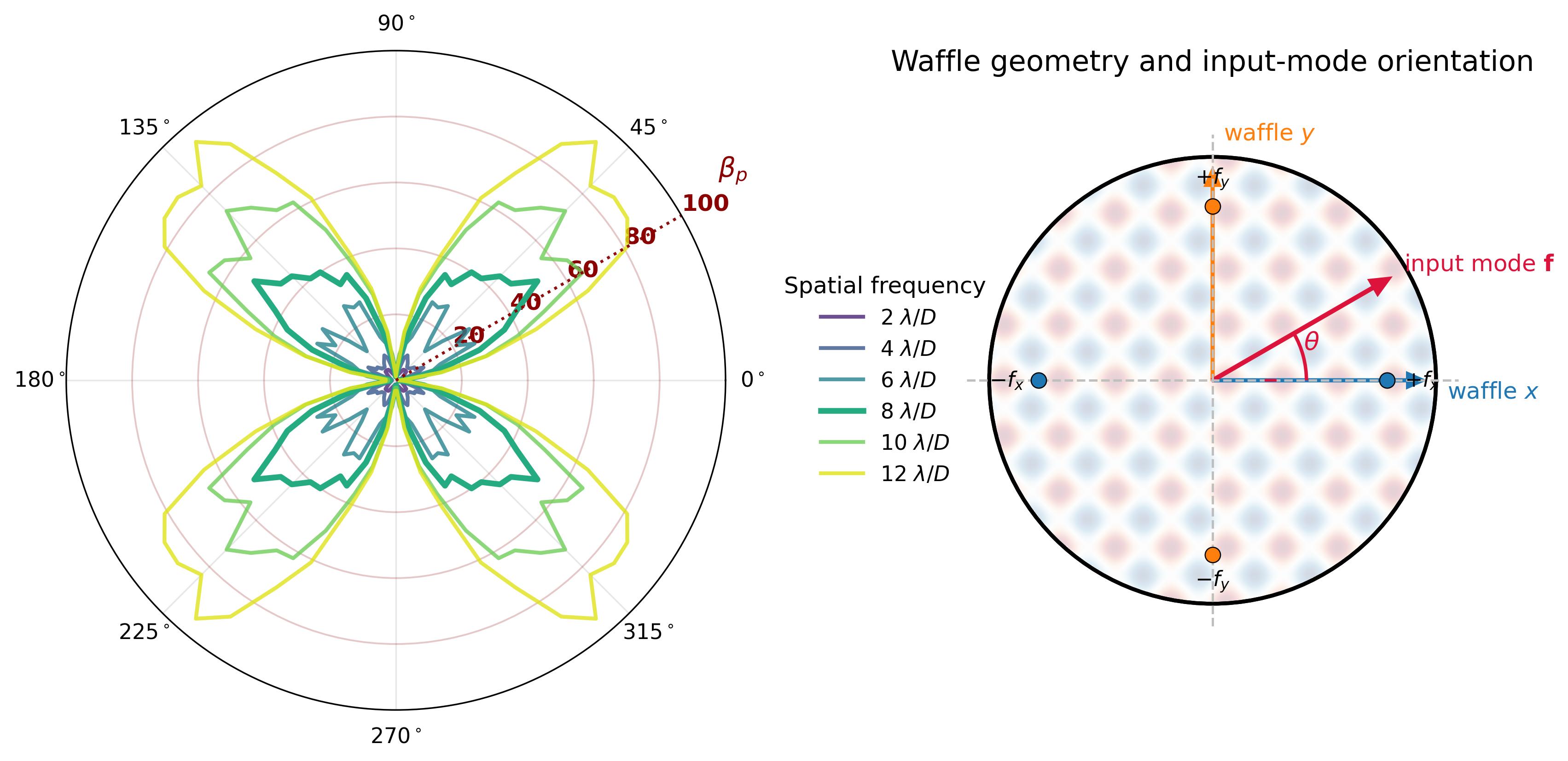}
    \caption{
    Angular dependence of the photon-noise-limited sensitivity coefficient $\beta_p$ for the four-reference phantom ZWFS at several Fourier spatial frequencies. Left: polar representation of $\beta_p$, showing the fourfold anisotropy imposed by the two orthogonal carrier axes. Lower $\beta_p$ corresponds to greater photon efficiency. Right: schematic of the four engineered satellite references and the orientation angle $\theta$ of the input Fourier mode relative to the $x$ carrier axis. The result demonstrates that the sensitivity of this particular reference geometry depends on both spatial frequency magnitude and orientation.
    }
    \label{fig:phantom_zwfs_ansiotropy}
\end{figure}

The resulting response, shown in Fig.~\ref{fig:phantom_zwfs_ansiotropy}, is strongly anisotropic. The lowest values of $\beta_p$ occur when the input Fourier perturbation is favourably aligned with the engineered reference geometry, whereas substantially poorer sensitivity is obtained for oblique orientations. The angular contrast becomes more pronounced at higher spatial frequencies. For example, at $4\,\lambda/D$ the densely sampled response spans approximately $\beta_p\simeq1$--$9$, with a median value of $5.28$, while at $8\,\lambda/D$ it spans approximately $\beta_p\simeq1$--$52$, with a median value of $27.02$.

This angular dependence is a direct consequence of the discrete reference geometry rather than an intrinsic property of the broader phantom ZWFS framework. The two orthogonal pupil carriers generate reference channels only along the corresponding focal plane directions, and therefore do not preserve the rotational symmetry of a conventional circular ZWFS phase mask. The Fourier sensitivity of this realization must consequently be written as
\begin{equation}
\beta_p
=
\beta_p(|\mathbf{f}|,\theta),
\end{equation}
rather than as a function of spatial frequency alone. Figure~\ref{fig:phantom_zwfs_ansiotropy} makes this structure explicit. The polar representation shows the fourfold symmetry imposed by the reference geometry, while the accompanying schematic defines the orientation of the input Fourier vector relative to the two carrier axes.

The anisotropy should not necessarily be interpreted as a defect of the engineered reference concept. Rather, it demonstrates that the geometry of the coherent references redistributes wavefront information before detection. For a general purpose wavefront sensor, additional carrier directions could in principle be introduced to make the response more angularly uniform, although the available reference power would then be distributed over a larger number of channels. Conversely, when the controlled aberration subspace or pupil geometry is known beforehand, a deliberately structured reference geometry may be desirable. The segmented-pupil example in Section~\ref{sec:segment_piston_example} illustrates the latter case by replacing the present two orthogonal carrier axes with three axes matched to the symmetry of the segmented aperture.

For the remainder of this section we retain the minimal four-reference geometry. It provides a simple test case in which the connection between reference placement and Fourier-space sensitivity can be examined directly before moving to application specific reference architectures.

\subsection{Sensitivity as a function of satellite separation and phase-mask diameter}
\label{sec:sensitivity_vs_design}

Having established the angular structure of the four-reference response, we next examine how its photon-noise-limited sensitivity varies with two geometric parameters: the satellite separation $\rho_{\mathrm{sat}}$ and the diameter $d_{\mathrm{mask}}$ of the localized satellite phase-shifting regions. This is a restricted characterization of the four-reference family, rather than a global optimization over all possible engineered reference geometries.

Throughout the scan we retain the phase progression
\begin{equation}
\left(
0,\frac{\pi}{2},\pi,\frac{3\pi}{2}
\right)
\end{equation}
for the four satellite references. This is the simplest symmetric quadrature choice: opposite phase shifts provide sign-reversed responses, while adjacent phase shifts sample orthogonal in-phase and quadrature projections. We do not claim that this phase set is globally optimal. More general phase assignments, carrier orientations, or numbers of reference channels represent additional design degrees of freedom beyond the restricted scan considered here. For each point on the $(\rho_{\mathrm{sat}},d_{\mathrm{mask}})$ grid, we evaluate $\beta_p$ for the six spatial frequencies $|\mathbf{f}| = \{2,4,6,8,10,12\}\,\lambda/D$ and three representative orientations, $\theta=\{0^\circ,45^\circ,90^\circ\}$. The $0^\circ$ and $90^\circ$ cases are aligned with the two carrier axes, while the $45^\circ$ case is equally inclined to both. These three directions do not exhaust the angular structure demonstrated in Section~\ref{sec:angular_anisotropy}, but provide a compact sampling of aligned and oblique responses for the present two-dimensional design scan.

\begin{figure}
    \centering
    \includegraphics[width=0.95\linewidth]{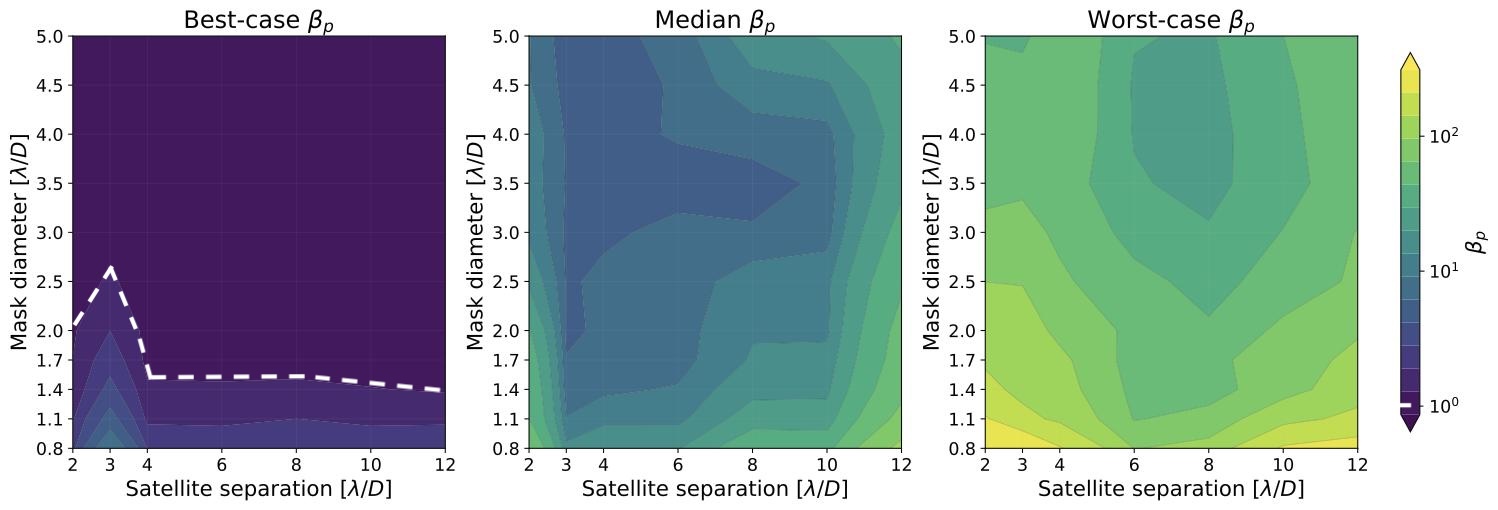}
    \caption{
    Restricted design scan for the four-reference phantom ZWFS using the photon-noise-limited sensitivity coefficient $\beta_p$. The panels show the sampled minimum, median, and maximum $\beta_p$ over different Fourier spatial frequencies and orientations, as functions of satellite separation $\rho_{\mathrm{sat}}$ and satellite phase-mask diameter $d_{\mathrm{mask}}$. These sampled quantities summarize the aligned and oblique responses used for the present design scan and should not be interpreted as exact extrema over all possible orientations. Lower values correspond to greater photon efficiency. The ideal classical-ZWFS reference $\beta_p\simeq1$ under the adopted normalization is indicated by the white dashed contour.
    }
    \label{fig:phantom_zwfs_design_scan}
\end{figure}

We summarize each design point by the minimum, median, and maximum $\beta_p$ over these 18 sampled Fourier perturbations. In the following we therefore refer to these quantities as the \emph{sampled} best-, median-, and worst case sensitivities. They should not be interpreted as exact extrema over all Fourier orientations. Figure~\ref{fig:phantom_zwfs_design_scan} shows the resulting maps. The sampled best case response approaches the ideal classical-ZWFS reference, $\beta_p\simeq1$ under the adopted Guyon normalization, once the satellite separation exceeds approximately $3\,\lambda/D$ and the phase-mask diameter exceeds approximately $1.5\,\lambda/D$. These values are better interpreted as transition scales within the restricted four-reference family than as unique optima. Beyond a separation of approximately $3\,\lambda/D$, the engineered satellite PSF copies are sufficiently separated from the central diffraction order and from neighbouring first-order references that further separation produces comparatively little improvement in the favourable orientations.

The preferred phase-mask diameter can be understood through the same interference visibility argument that motivates the conventional ZWFS dimple size. A classical $1.06\,\lambda/D$ phase dimple contains approximately half of the diffraction-limited PSF energy, producing a near balanced phase-shifted and unshifted field and therefore high interference visibility \citep[e.g.,][]{ndiaye_2013_zelda}. The photon balance is different in the core preserving phantom realization because most of the zeroth diffraction order is first removed through science coupling.

For the adopted waffle amplitude $a=0.5$ rad, the exact Jacobi--Anger expansion gives approximately $77.6\%$ of the incident power in the zeroth diffraction order and approximately $5.2\%$ in each of the four first-order satellites. If approximately $78\%$ of the zeroth-order power is coupled into the science fibre, the uncoupled zeroth-order contribution remaining in the sensing branch contains approximately $17\%$ of the incident power, while the four first-order satellites together contain approximately $21\%$. A phase-shifting diameter of $3.5\,\lambda/D$ contains approximately $88\%$ of each satellite PSF, corresponding to approximately $18\%$ of the incident power in the phase-shifted satellite regions. Neglecting the smaller higher diffraction orders, the remaining unshifted sensing field contains approximately $20\%$, giving a near balanced phase-shifted/unshifted photon distribution. This provides a simple visibility-based explanation for why the favourable satellite phase-mask diameter in this core preserving realization is larger than the conventional $1.06\,\lambda/D$ ZWFS dimple.

The median and sampled worst case maps show a different aspect of the same design trade. A geometry that provides near classical sensitivity for favourably aligned Fourier modes can remain substantially less efficient for oblique modes because of the directional structure demonstrated in Fig.~\ref{fig:phantom_zwfs_ansiotropy}. For the baseline modal benchmark in Section~\ref{sec:benchmark_classical_vs_phantom}, we therefore select a configuration from the broad region providing favourable median sensitivity while maintaining clear spatial separation of the engineered references, rather than treating the minimum-$\beta_p$ location as a unique optimum.

More generally, the results of this section demonstrate that engineered reference geometry changes the distribution of wavefront information across spatial frequency and angular directions. In the four-reference realization this appears as strong fourfold anisotropy and a trade between satellite separation, mask diameter, and photon efficiency. The following benchmark asks how this structured sensitivity translates into modal reconstruction relative to a classical ZWFS. Later, in Section~\ref{sec:segment_piston_example}, the reference geometry is instead selected around a known segmented-pupil wavefront subspace, illustrating that angularly averaged Fourier sensitivity need not be the appropriate design objective for every application.

\section{Modal reconstruction benchmark of the four-reference realization}
\label{sec:benchmark_classical_vs_phantom}

The Fourier analysis in Section~\ref{sec:four_reference_characterization} characterizes the small signal response of the four-reference phantom ZWFS independently of a particular reconstruction basis. We now ask how this structured sensitivity appears in a conventional calibrated modal reconstruction. To do so, we compare the same standalone four-reference realization with a classical Zernike wavefront sensor under a common scalar propagation and reconstruction framework.

The purpose of this benchmark is to provide a controlled example of how a fixed engineered reference geometry redistributes modal response, linearity, and cross-talk relative to a rotationally symmetric classical ZWFS. This provides a bridge between the Fourier-space characterization above and the application specific reference design considered later in the paper.

\subsection{Simulation setup}

Both sensors are simulated using the same discretized pupil and focal plane grids. The input pupil is a circular aperture sampled on a $128\times128$ grid, while focal plane propagation is performed on a zero padded $512\times512$ grid. Wavefronts are expanded in a Zernike basis excluding piston, and the first 16 modes are retained for calibration and reconstruction. Each mode is normalized to unit RMS over the illuminated pupil. For the classical ZWFS, the focal plane mask is a circular phase-shifting region of diameter $1.06\,\lambda/D$ with a phase shift of $\pi/2$. For the four-reference phantom ZWFS, the central phase mask is absent, $M_0=1$, and the core preserving photonic routing described in Section~\ref{sec:photonic_realization} is retained. A pupil modulation of amplitude $a=0.5$ rad at 7 cycles per pupil generates four first-order satellite references at a separation of $7\,\lambda/D$. These references are phase shifted by $\left( 0,\frac{\pi}{2},\pi,\frac{3\pi}{2}\right),$ using circular focal plane regions of diameter $3.5\,\lambda/D$.

This configuration is chosen as a representative point within the broad favourable region identified in Fig.~\ref{fig:phantom_zwfs_design_scan}, while placing the satellite references sufficiently far from the central diffraction order and from one another for clean carrier isolation. The $3.5\,\lambda/D$ phase-mask diameter is also consistent with the approximate phase-shifted/unshifted photon balance discussed in Section~\ref{sec:sensitivity_vs_design}. The configuration should therefore be interpreted as a convenient benchmark realization rather than as a globally optimal phantom geometry.

The adopted waffle amplitude places a useful fraction of the incident light into the first-order reference channels. In the exact Jacobi--Anger expansion, the first-order satellite field scales as $J_1(a)J_0(a)$; for $a=0.5$ rad this is approximately $0.23$, corresponding to approximately $5\%$ of the optical power in each first-order satellite. Although the weak waffle expansion of Section~\ref{sec:weak_waffle} is useful for interpreting the carrier structure, the numerical propagation used for this benchmark retains the full exponential pupil modulation.

The phantom measurement vector is the carrier-only observable defined in Eq.~(\ref{eq:y_sat}),
\begin{equation}
\mathbf{y}_{\rm ph}
=
\begin{bmatrix}
\Re(M_x)\\
\Im(M_x)\\
\Re(M_y)\\
\Im(M_y)
\end{bmatrix}_{P>0},
\end{equation}
formed from the differential pupil intensity relative to the zero aberration reference. The $x$ and $y$ carrier components are mixed to baseband and low-pass filtered as described in Section~\ref{sec:theory_reconstruction}. For the classical ZWFS, the measurement vector is instead the differential pupil intensity itself,
\begin{equation}
\mathbf{y}_{\rm Z}
=
\left[
\Delta I_{\rm Z}(\mathbf r)
\right]_{P>0}.
\end{equation}

For both sensors, the zero aberration reference is treated as known exactly and the interaction matrix is constructed using the push--pull procedure of Section~\ref{sec:build_interaction_matrix}, with a calibration amplitude of $\epsilon=0.1$ rad RMS. The modal reconstructor is the corresponding truncated Moore--Penrose pseudo-inverse, and residual phase RMS is evaluated over the illuminated pupil. Where a photon normalization is required, the comparison uses an equal detected WFS-channel photon convention,
\begin{equation}
N_{\rm ph,WFS}^{\rm pZ}
=
N_{\rm ph,WFS}^{\rm Z}.
\end{equation}
This isolates differences in the optical encoding and reconstruction from the additional photon-sharing penalty associated with coupling the central core into a science channel. A comparison at fixed incident stellar flux would additionally require
\begin{equation}
N_{\rm ph,WFS}^{\rm pZ}
=
\eta_{\rm refl}
N_{\rm ph,inc},
\end{equation}
as discussed in Section~\ref{sec:sensitivity_metric}. For the adopted $a=0.5$ configuration, approximately $J_0(0.5)^4\simeq0.78$ of the light remains in the central diffraction order. Combining this with an ideal diffraction-limited single mode coupling of $\eta_{\rm coup,0}\simeq0.78$ gives an approximate returned fraction
\begin{equation}
\eta_{\rm refl}
\simeq
1-0.78\times0.78
\simeq
0.39,
\end{equation}
before additional optical losses. The main numerical parameters are summarized in Table~\ref{tab:phantom_benchmark_params}.

\begin{table}[t]
    \centering
    \caption{Main parameters adopted for the modal reconstruction benchmark of the classical ZWFS and the four-reference phantom realization.}
    \label{tab:phantom_benchmark_params}
    \begin{tabular}{lll}
        \hline
        Quantity & Symbol / description & Value \\
        \hline
        Pupil sampling
            & $N_{\mathrm{pupil}}$
            & $128\times128$ \\
        Focal FFT sampling
            & $N_{\mathrm{FFT}}$
            & $512\times512$ \\
        Normalized pupil radius
            & $R_{\mathrm{pupil}}$
            & 0.32 \\
        Number of reconstructed modes
            & $N_{\mathrm{modes}}$
            & 16 \\
        Modal basis
            &
            & Zernike, piston excluded \\
        Calibration amplitude
            & $\epsilon$
            & 0.1 rad RMS \\
        Modal amplitude scan
            & $\alpha_{\rm inj}$
            & $-1$ to $+1$ rad RMS \\
        Number of scan points
            &
            & 31 \\
        Classical mask diameter
            &
            & $1.06\,\lambda/D$ \\
        Classical phase shift
            &
            & $\pi/2$ \\
        Phantom waffle amplitude
            & $a$
            & 0.50 rad \\
        Phantom carrier frequency
            & $f$
            & 7 cycles per pupil \\
        Phantom satellite separation
            & $\rho_{\rm sat}$
            & $7.0\,\lambda/D$ \\
        Phantom satellite mask diameter
            & $d_{\rm mask}$
            & $3.5\,\lambda/D$ \\
        Phantom satellite phase shifts
            &
            & $(0,\pi/2,\pi,3\pi/2)$ \\
        WFS photon budget
            & $N_{\mathrm{ph,WFS}}$
            & $5\times10^8$ \\
        Photon comparison convention
            &
            & equal detected WFS photons \\
        \hline
    \end{tabular}
\end{table}

\subsection{Modal response and reconstruction error}
\label{sec:benchmark_procedure}

\begin{figure}
    \centering
    \includegraphics[width=0.8\linewidth]{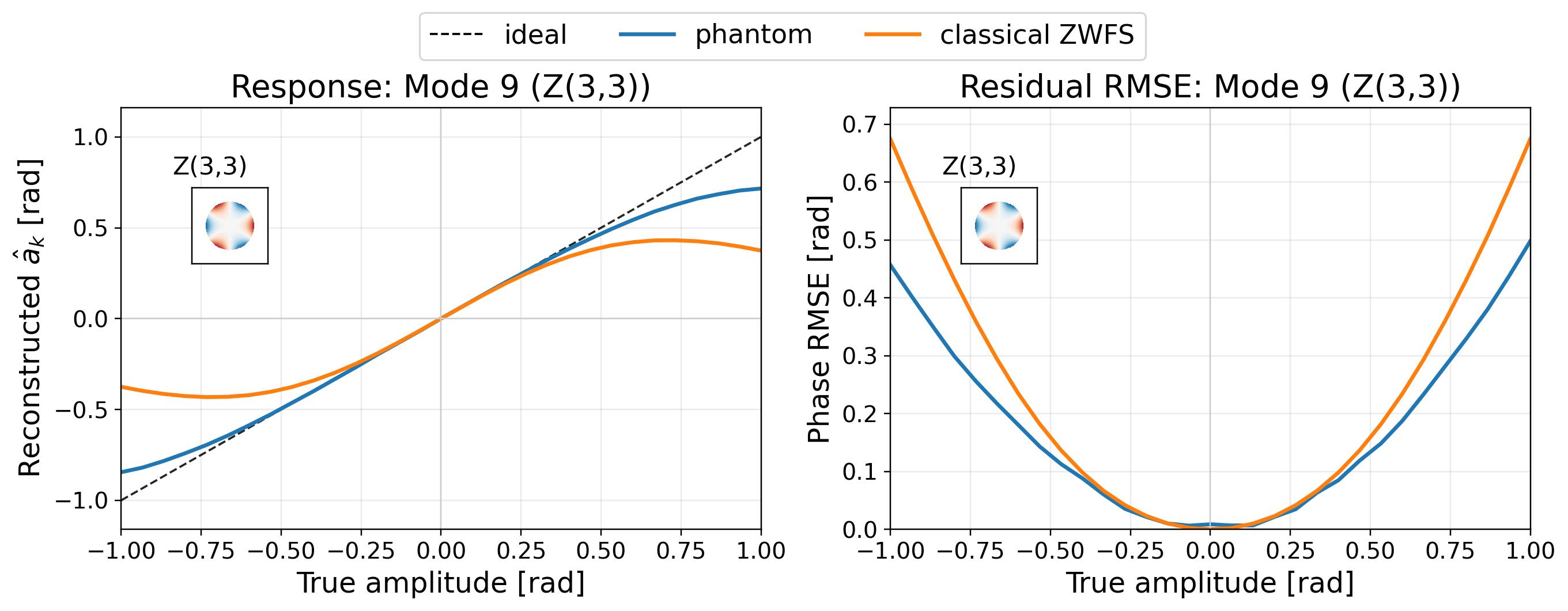}
    \includegraphics[width=0.8\linewidth]{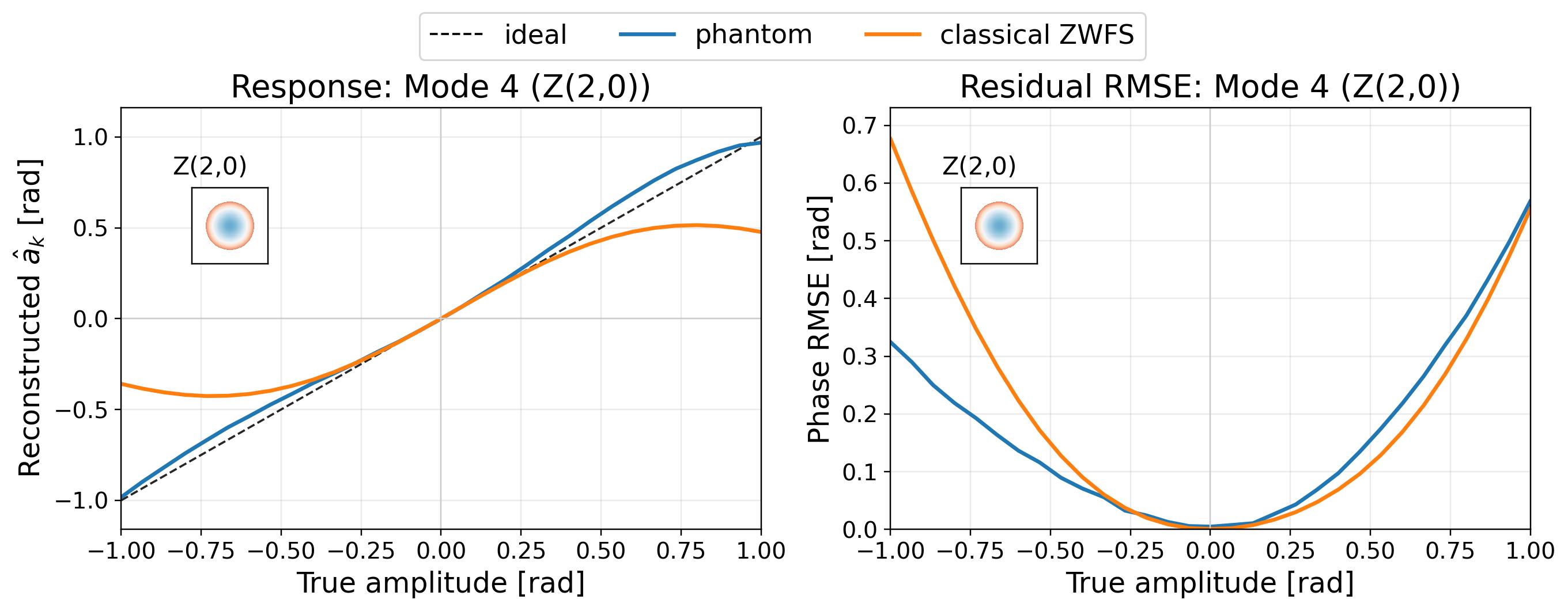}
    \caption{
    Representative modal reconstruction benchmarks for the four-reference phantom ZWFS and a classical ZWFS. Top: oblique trefoil, $Z(3,3)$, for which the four-reference phantom realization gives a reconstructed coefficient closer to the ideal linear response and a lower residual phase RMS over much of the tested amplitude range. Bottom: defocus, $Z(2,0)$, for which the response remains
    closer to the injected coefficient over an extended range. This produces a
    lower residual phase RMS on the negative amplitude side, whereas at positive
    amplitudes the improved response yields little reduction in total
    residual error because of increased modal cross-talk. These examples illustrate that the effect of the engineered reference geometry is mode dependent rather than a uniform increase in sensing performance.
    }
    \label{fig:benchmark_examples}
\end{figure}

To isolate the mode dependence of the nonlinear sensor response, each Zernike mode is injected individually over amplitudes from $-1$ to $+1$ rad RMS while reconstruction is performed over the complete 16 mode basis. For every injected amplitude we record (i) the reconstructed coefficient of the injected mode, (ii) the total residual phase RMS after reconstruction, and (iii) the full reconstructed coefficient vector. The first quantity measures the self-response of the mode, while the latter two reveal whether an apparently extended modal response is accompanied by leakage into other reconstructed modes. Representative cases are shown in Fig.~\ref{fig:benchmark_examples}. The complete modal response and residual error curves are given in Figs.~\ref{fig:appendix_all_mode_response} and~\ref{fig:appendix_all_mode_rmse}, and the corresponding cross-talk matrices are shown in Fig.~\ref{fig:appendix_crosstalk}.

For the oblique trefoil mode $Z(3,3)$, the four-reference phantom ZWFS produces a reconstructed coefficient that remains closer to the ideal linear response over much of the tested amplitude range than the classical ZWFS. This is accompanied by a lower total residual phase RMS over much of the same range. The response is nevertheless asymmetric about zero amplitude, showing that the improvement is specific to the nonlinear response produced by this particular reference geometry rather than a simple increase in a symmetric capture range.

Defocus, $Z(2,0)$, provides a complementary example. The reconstructed phantom coefficient remains substantially closer to the injected value toward both ends of the tested amplitude range than for the classical ZWFS. On the negative amplitude side, this improved self-response is accompanied by a lower total residual phase RMS. On the positive amplitude side, however, the improved self-response produces little corresponding reduction in the total residual phase RMS. The cross-talk matrices in Fig.~\ref{fig:appendix_crosstalk} show that leakage into other reconstructed modes offsets part of the improved defocus response. Thus a more linear response of the injected coefficient does not necessarily translate into a comparable improvement in reconstruction of the full phase.

Across the complete modal basis there is no uniform ordering between the two sensors. Some modes remain more favourably reconstructed by the classical ZWFS, whereas others show a broader or more nearly linear response with the four-reference phantom geometry. This is consistent with the Fourier-space results of Section~\ref{sec:four_reference_characterization}: introducing a discrete engineered reference geometry redistributes the available phase information rather than producing a uniform improvement over all wavefront directions.

The significance of this benchmark is therefore not the particular performance of the four-reference configuration itself. Rather, it demonstrates in a conventional modal basis that the optical reference geometry changes both the desired modal response and the nonlinear cross-talk structure. A general purpose sensor would normally seek a suitable compromise across the controlled basis. If the important wavefront subspace is known in advance, however, the reference geometry can instead be designed around that subspace. The segment-piston application in Section~\ref{sec:segment_piston_example} develops this second approach explicitly by replacing the present four-reference geometry with a carrier pattern matched to the symmetry and differential piston modes of a segmented pupil.

\section{Application specific reference design: hybrid segment-piston sensing}
\label{sec:segment_piston_example}
Segment phasing is a well established application of Zernike phase-contrast sensing. Zernike based sensors have been demonstrated on-sky for segmented mirror phasing \citep{Surdej2010,Vigan2011}, and have subsequently been
developed for diffraction-limited fine cophasing of segmented apertures \citep{JaninPotiron2017}. The problem remains directly relevant to current extremely large segmented telescopes, including the ELT, where Zernike-based phasing concepts continue to be developed \citep{Kellerer2026}.

The four-reference realization considered in Sections~ \ref{sec:four_reference_characterization} and \ref{sec:benchmark_classical_vs_phantom} was deliberately simple and was characterized over a general Fourier and Zernike mode space. Segment piston provides a more application-specific test because both the controlled wavefront subspace and the pupil geometry are known in advance. A conventional ZWFS provides high sensitivity close to the phased state, but the mapping from segment piston to pupil intensity becomes increasingly nonlinear and poorly conditioned as the phase excursion grows. We therefore ask whether engineered off-axis references can be added specifically to improve the observability of this differential-piston subspace while retaining the conventional central ZWFS response.

\begin{figure}
    \centering
    \includegraphics[width=0.98\linewidth]
    {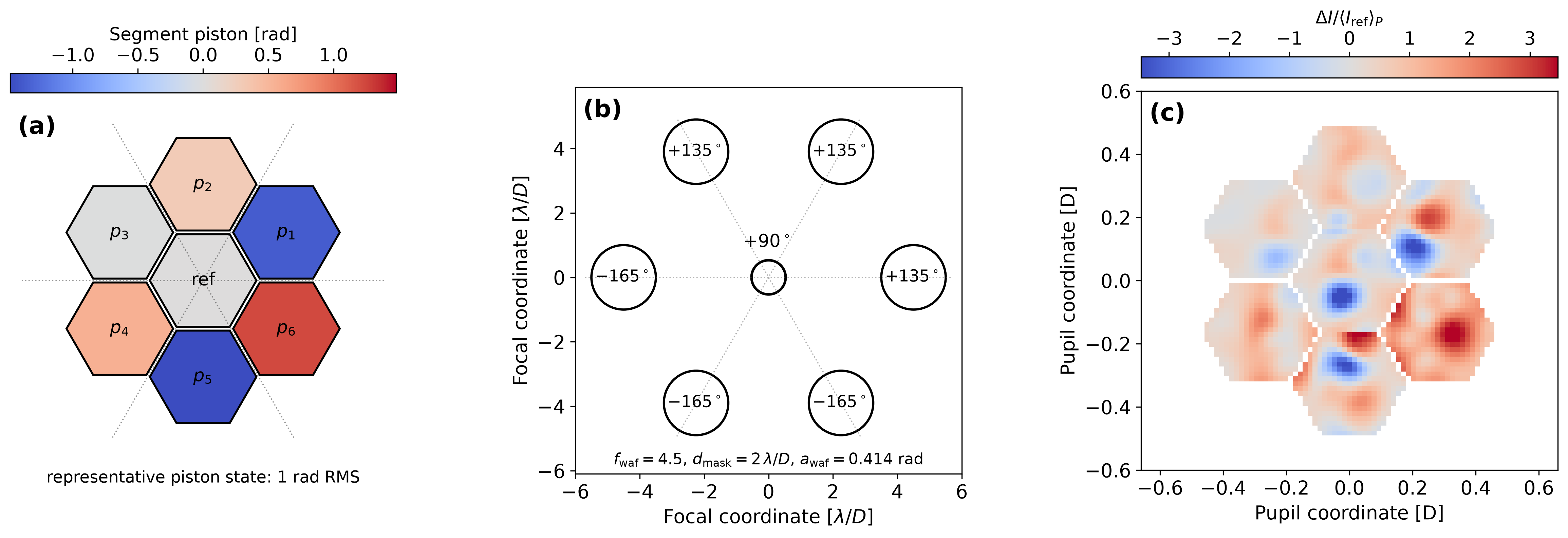}
    \caption{
    Representative hybrid ZWFS-phantom measurement for differential
    segment piston. 
    (a) Example seven-segment piston state with 1~rad RMS simultaneous piston;
    the central segment defines the zero-piston reference and the six outer
    segments carry the differential piston values.
    (b) Corresponding focal plane phase-mask geometry. A conventional
    $1.06\,\lambda/D$ central ZWFS region applies a $\pi/2$ phase shift, while
    three pupil carriers generate six satellite references at
    $f_{\rm waf}=4.5$ cycles per pupil. The satellite phase-mask diameter is
    $2.0\,\lambda/D$, the waffle amplitude is $a_{\rm waf}=0.414$ rad, and the
    positive- and negative frequency satellite groups are assigned phase shifts
    of $135^\circ$ and $-165^\circ$, respectively.
    (c) Differential hybrid pupil intensity produced by the same piston state,
    shown as
    $\Delta I_{\rm hyb}/\langle I_{\rm ref}\rangle_{P}$, where
    $\Delta I_{\rm hyb}=I_{\rm hyb}(\mathbf p)-I_{\rm hyb}(\mathbf 0)$ and
    $\langle I_{\rm ref}\rangle_{P}$ is the mean reference intensity over the
    illuminated pupil. The measured pupil image contains both the conventional
    central ZWFS response and the spatial structure introduced by the engineered
    satellite references.
    }
    \label{fig:segment_hybrid_architecture}
\end{figure}

We consider the seven-segment pupil shown in Fig.~\ref{fig:segment_hybrid_architecture}(a). The central segment defines the
phase reference, $p_0=0$, and the six outer segments carry independent
differential pistons,
\begin{equation}
    \phi(\mathbf r)
    =
    \sum_{s=1}^{6}
    p_s S_s(\mathbf r),
    \label{eq:segment_piston_phase}
\end{equation}
where $S_s(\mathbf r)$ is the support of segment $s$. Three sensors are compared using the same segmented pupil. The scalar ZWFS uses a single central focal plane phase region of diameter $1.06\,\lambda/D$ with a phase shift of $+\pi/2$. The vector ZWFS benchmark uses two complementary central phase-contrast outputs with phase shifts $+\pi/2$ and $-\pi/2$, with the total detected WFS photon budget divided equally between the two outputs \citep{doelman_2019_vzwfs}. The hybrid sensor retains the same $+\pi/2$ central ZWFS region as the scalar sensor and adds engineered phantom references around it.

For the hybrid sensor, three pupil plane carrier axes separated by $60^\circ$ are introduced,
\begin{equation}
    \phi_{\rm waf}(\mathbf r)
    =
    a_{\rm waf}
    \sum_{j=1}^{3}
    \sin\left(
        2\pi f_{\rm waf}
        \hat{\mathbf e}_j\cdot\mathbf r
    \right),
    \label{eq:segment_waffle}
\end{equation}
where the carrier directions are separated by the symmetry of the segmented aperture. The modulation generates six first-order satellite references at $\pm f_{\rm waf}\hat{\mathbf e}_j$. Circular phase-shifting regions are placed on these satellites in addition to the central ZWFS phase region, as shown in Fig.~\ref{fig:segment_hybrid_architecture}(b).

For the representative configuration used here, the carrier frequency and satellite phase-mask diameter were first fixed at $f_{\rm waf}=4.5$ cycles per pupil and $d_{\rm mask}=2.0\,\lambda/D$ from a preceding coarse symmetry preserving geometry scan. We then optimized the auxiliary-reference strength and focal plane phase shifts while retaining the three carrier axes at $0^\circ$, $60^\circ$, and $120^\circ$.

Candidate designs were evaluated from the detector level Fisher information over independent random differential piston direction sets. The performance metric combined four quantities: the local Cram\'er--Rao lower bound at $0.2$ rad RMS piston, normalized to the scalar ZWFS; the Cram\'er--Rao lower bound over $0.8$, $1.2$, and $1.5$ rad RMS piston, normalized to the vector ZWFS; the corresponding high-piston Fisher-matrix condition number; and, with smaller weight, the conditioning of the exact harmonic response. The respective weights were $0.30$, $0.40$, $0.20$, and $0.10$. Independent optimization and validation piston-direction banks were used during this search, and the final reconstruction benchmark uses a separate set of 200 random piston directions.

A subsequent refinement extended the waffle amplitude to $2$ rad and searched the symmetry preserving conjugate phase shift family with phase offsets $\{-15^\circ,0^\circ,+15^\circ\}$ and phase separations from $110^\circ$ to $160^\circ$ in $5^\circ$ steps. Rather than selecting the absolute information optimum, we retained all designs with a performance score within $\Delta S=0.02$ of the minimum and selected from this set the configuration preserving the largest fraction of power in the central ZWFS core. This gives the balanced configuration used below: $a_{\rm waf}=0.414$ rad, with phase shifts of $+135^\circ$ on the three positive frequency references and $-165^\circ$ on the three negative frequency references. Approximately $77\%$ of the unmodulated central core power is retained, while approximately $17\%$ of the incident power lies within the six satellite phase-shifting regions. This configuration is therefore a deliberately balanced representative design.

\subsection{Hybrid measurement}

For a piston vector $\mathbf p$, we define the hybrid differential pupil image
relative to the phased reference as
\begin{equation}
    \Delta I_{\rm hyb}(\mathbf r;\mathbf p)
    =
    I_{\rm hyb}(\mathbf r;\mathbf p)
    -
    I_{\rm hyb}(\mathbf r;\mathbf 0).
    \label{eq:segment_deltaI}
\end{equation}
This is the directly measured pupil plane signal; it contains both the conventional central ZWFS response and the intensity modulation introduced by the six auxiliary references. The three complex carrier observables are obtained from this same detector image using the demodulation introduced in Section~ \ref{sec:theory_reconstruction},
\begin{equation}
    M_j(\mathbf r)
    =
    \mathrm{LPF}
    \left[
        \Delta I_{\rm hyb}(\mathbf r)
        \exp\left(
            -i2\pi f_{\rm waf}
            \hat{\mathbf e}_j\cdot\mathbf r
        \right)
    \right],
    \qquad j=1,2,3.
    \label{eq:segment_carrier_demod}
\end{equation}
The hybrid feature vector is therefore
\begin{equation}
    \mathbf y_{\rm hyb}
    =
    \begin{bmatrix}
        \Delta I_{\rm hyb}\\
        \Re(M_1)\\
        \Im(M_1)\\
        \Re(M_2)\\
        \Im(M_2)\\
        \Re(M_3)\\
        \Im(M_3)
    \end{bmatrix}_{P>0}.
    \label{eq:segment_hybrid_features}
\end{equation}
The direct pupil image and the demodulated carrier maps are not independent photon measurements: all seven feature blocks are deterministic transforms of the same detected intensity. Their noise is consequently correlated, and the full feature covariance is retained in the noise-weighted reconstruction. This construction is an explicit realization of the general hybrid measurement described in Eq.~(\ref{eq:y_hybrid_general}). The satellite
channels augment rather than replace the central phase-contrast response.

\subsection{Exact harmonic model for segment piston}

Segment piston is intrinsically periodic, so a local polynomial expansion is not the natural forward model for this application. Instead, the
monochromatic piston dependence can be written exactly in a finite harmonic basis. Let $E_s(\mathbf r)$ denote the detector-plane complex field produced by segment $s$ with unit complex amplitude after propagation through all fixed optics of the sensor. The total field for piston vector $\mathbf p$ is then
\begin{equation}
    E(\mathbf r;\mathbf p)
    =
    \sum_{s=0}^{6}
    E_s(\mathbf r)e^{ip_s},
    \qquad p_0=0.
    \label{eq:segment_field_sum}
\end{equation}
The corresponding intensity is
\begin{equation}
    I(\mathbf r;\mathbf p)
    =
    \sum_{s=0}^{6}|E_s|^2
    +
    2
    \sum_{s<t}
    \Re\left[
        E_s^*(\mathbf r)
        E_t(\mathbf r)
        e^{i(p_t-p_s)}
    \right].
    \label{eq:segment_intensity_pairwise}
\end{equation}
Subtracting the phased reference, $\mathbf p=\mathbf 0$, removes the
piston independent terms. If
\begin{equation}
    E_s^*E_t
    =
    A_{st}+iB_{st},
\end{equation}
each segment pair contributes only the two real harmonic coordinates
\begin{equation}
    \sin(p_t-p_s),
    \qquad
    1-\cos(p_t-p_s).
\end{equation}
For the six central-to-outer pairs these reduce to $\sin p_s$ and $1-\cos p_s$, while the 15 outer-to-outer pairs depend on the corresponding piston differences. There are therefore
\begin{equation}
    2{7\choose2}=42
\end{equation}
real harmonic coordinates in total. Because formation of the direct pupil feature and the carrier demodulation are linear operations on the differential intensity, the same harmonic coordinates describe any of the sensor feature vectors considered here. We therefore write
\begin{equation}
    \mathbf y(\mathbf p)
    =
    \mathbf H_{\rm harm}
    \mathbf q(\mathbf p),
    \label{eq:segment_harmonic_model}
\end{equation}
where $\mathbf q(\mathbf p)$ contains the 42 sine and cosine difference terms defined above and $\mathbf H_{\rm harm}$ contains their corresponding calibrated measurement space response vectors. Within the adopted monochromatic scalar optical model, Eq.~(\ref{eq:segment_harmonic_model}) is an exact representation of the piston dependence rather than a truncated Taylor expansion. The piston estimate is obtained from a noise-weighted nonlinear fit,
\begin{equation}
    \hat{\mathbf p}
    =
    \arg\min_{\mathbf p}
    \left[
        \mathbf y_{\rm meas}
        -
        \mathbf H_{\rm harm}\mathbf q(\mathbf p)
    \right]^{\mathsf T}
    \mathbf C_n^{-1}
    \left[
        \mathbf y_{\rm meas}
        -
        \mathbf H_{\rm harm}\mathbf q(\mathbf p)
    \right].
    \label{eq:segment_piston_estimator}
\end{equation}
which corresponds to the usual generalized least-squares form for a
measurement with covariance $\mathbf C_n$
\citep{Kay1993}. The same harmonic model and fitting procedure are applied independently to the classic scalar ZWFS, vector ZWFS, and phantom ZWFS sensor.

To avoid giving the estimator prior knowledge of the piston amplitude, the final benchmark uses one fixed sensor specific covariance matrix $\mathbf C_n$, estimated from an independent calibration ensemble spanning the tested piston range. This covariance is calculated once for each sensor and is then held fixed for all injected piston amplitudes and all final test realizations. For the hybrid measurement, $\mathbf C_n$ includes the correlations between the direct pupil image and its three demodulated carrier maps.

\subsection{Reconstruction performance}

The reconstruction comparison uses 200 fixed random piston directions, with the simultaneous piston RMS varied between $0.1$ and $1.5$ rad. Individual segment pistons are restricted to
\begin{equation}
    |p_s|<0.95\pi,
\end{equation}
so that all test states remain within the principal monochromatic phase branch. Poisson noise is applied using a total detected WFS photon budget of $10^5$ photons per measurement. The same total detected photon number is used for all three sensors.

Figure~\ref{fig:segment_hybrid_performance}(a) shows the resulting wrapped piston reconstruction error. Close to the phased state, the hybrid and vector ZWFS provide similar precision. At $0.1$ rad RMS simultaneous piston, the median reconstruction errors are $0.0116$ and $0.0102$ rad RMS for the hybrid and vector ZWFS, respectively, with corresponding 90th percentile errors of $0.0172$ and $0.0167$ rad RMS.

The difference becomes increasingly favourable to the hybrid architecture as the piston excursion grows. At $1.5$ rad RMS, the hybrid gives a median reconstruction error of $0.0185$ rad RMS and a 90th percentile error of $0.0283$ rad RMS, compared with $0.0318$ and $0.0528$ rad RMS for the vector ZWFS. The engineered references therefore reduce the median and 90th percentile reconstruction errors by approximately $42\%$ and $46\%$, respectively, at the largest tested piston amplitude.

As a robustness test, we repeated the reconstruction using a covariance estimated separately at each injected piston RMS. The state matched and fixed global covariance results remain closely consistent, with differences of less than approximately $6\%$ in the median and 90th percentile errors for the final hybrid configuration over the complete tested range. The performance ordering is unchanged. We therefore use the fixed-covariance results throughout Fig.~\ref{fig:segment_hybrid_performance}(a), since this estimator does not require prior knowledge of the true piston amplitude.

\subsection{Detector level Fisher information}

The reconstruction result depends on the chosen nonlinear estimator. To test whether the hybrid advantage is already present in the optical measurement, we therefore also evaluate the local Fisher information directly from the detector intensities, before carrier demodulation or piston fitting. Under the equal detected WFS photon convention, we isolate the information contained in the spatial distribution of detected light. Let $I_i(\mathbf p)$ be the expected detector intensity in pixel $i$ within the common analysis support and define the normalized spatial distribution
\begin{equation}
    \pi_i(\mathbf p)
    =
    \frac{I_i(\mathbf p)}
         {\sum_j I_j(\mathbf p)}.
\end{equation}
For $N_{\rm sup}$ expected photons within that support, the conditional Poisson Fisher information matrix \citep{Fisher1922,Kay1993} for the six pistons is
\begin{equation}
    F_{st}
    =
    N_{\rm sup}
    \sum_i
    \frac{1}{\pi_i}
    \frac{\partial\pi_i}{\partial p_s}
    \frac{\partial\pi_i}{\partial p_t}.
    \label{eq:segment_fisher}
\end{equation}
The intensity derivatives are evaluated directly from the propagated complex fields. Full numerical details, including treatment of near-zero-intensity pixels and the common photon normalization, are given in Appendix~\ref{app:segment_fisher}.

\begin{figure}
    \centering
    \includegraphics[width=0.98\linewidth]
    {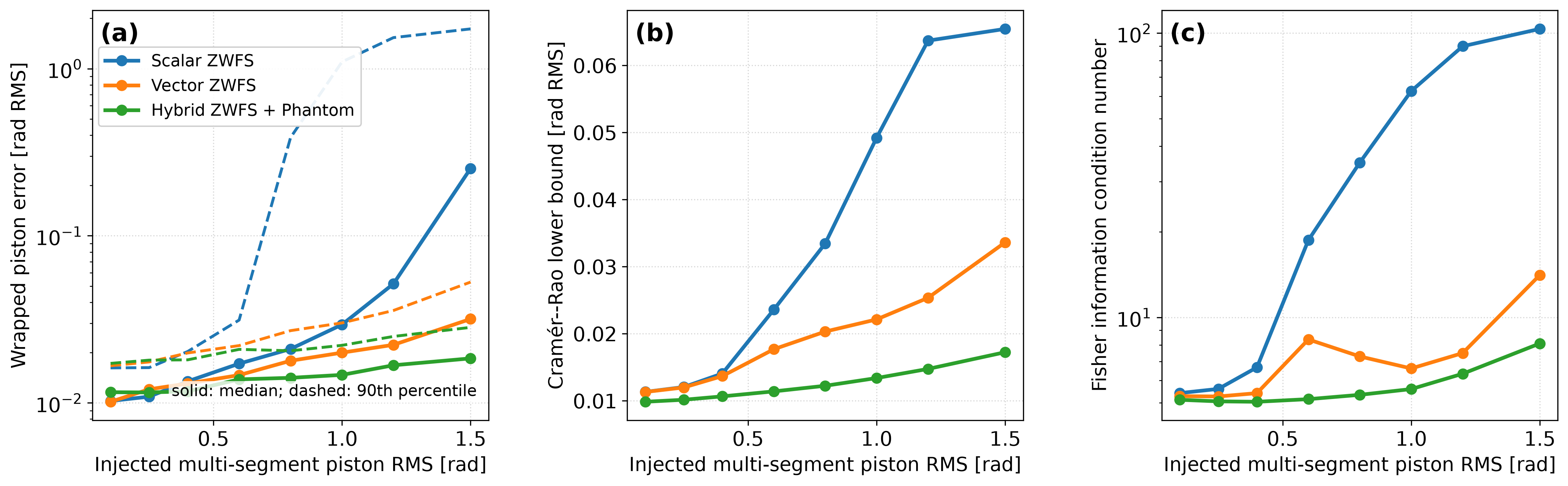}
    \caption{
    differential piston performance for the scalar ZWFS, vector ZWFS, and hybrid ZWFS with six auxiliary phantom references, using $10^5$ total
    detected WFS photons per measurement. (a) Wrapped piston reconstruction error for simultaneous random segment piston. Solid curves show the median and dashed curves the 90th percentile over 200 fixed test directions. (b) Local conditional Poisson Cram\'er--Rao lower bound calculated directly from the detector intensity distribution and its derivatives with respect to the six segment pistons. (c) Condition number of the corresponding Fisher information matrix. The hybrid architecture retains the small-piston precision of the conventional phase-contrast measurement while providing a better conditioned piston observable at larger phase excursions.
    }
    \label{fig:segment_hybrid_performance}
\end{figure}

For a six-dimensional piston vector, we summarize the local Cram\'er--Rao lower bound \citep{Rao1945,Cramer1946,Kay1993} by
\begin{equation}
    \sigma_{\rm CRLB}
    =
    \sqrt{
        \frac{1}{6}
        \operatorname{tr}
        \left(
            \mathbf F^{+}
        \right)
    },
    \label{eq:segment_crlb}
\end{equation}
where $\mathbf F^{+}$ denotes the Moore--Penrose inverse. We also calculate the condition number of the Fisher information matrix,
\begin{equation}
    \kappa_F
    =
    \frac{\lambda_{\max}(\mathbf F)}
         {\lambda_{\min}(\mathbf F)},
    \label{eq:segment_fisher_condition}
\end{equation}
where $\lambda_{\max}(\mathbf F)$ and $\lambda_{\min}(\mathbf F)$ are the largest and smallest eigenvalues of the Fisher information matrix, respectively. These eigenvalues quantify the local information along the principal directions of the six-dimensional piston parameter space. Thus $\kappa_F=1$ corresponds to equal information along all principal directions, whereas a large $\kappa_F$ indicates that at least one piston combination is substantially less constrained than the best measured combination. A smaller
$\kappa_F$ therefore indicates a better-conditioned six-parameter measurement.
Figures~\ref{fig:segment_hybrid_performance}(b) and \ref{fig:segment_hybrid_performance}(c) show that the engineered references change the detector level information content itself. At $1.5$ rad RMS simultaneous piston, the median local Cram\'er--Rao lower bound is approximately $0.0172$ rad RMS for the hybrid architecture compared with $0.0336$ rad RMS for the vector ZWFS. The corresponding median condition numbers of the Fisher information matrix are approximately $8.1$ and $14.1$, respectively. The scalar ZWFS becomes still more poorly conditioned over this regime.

The Fisher calculation is independent of the carrier demodulation and nonlinear estimator and therefore provides the most direct evidence that the additional references improve the optical encoding of the selected piston subspace. The Cram\'er--Rao lower bound is nevertheless a local precision metric: it does not describe global capture range and does not remove the periodic $2\pi$ ambiguity of monochromatic piston sensing.

Taken together, this example demonstrates a complementary use of the phantom architecture to the core preserving realization studied earlier. Rather than moving the phase-contrast reference away from the central PSF core, the engineered channels are added to an existing ZWFS and designed around a known wavefront subspace. In this case the three-axis geometry provides additional piston diversity while retaining the central ZWFS response. The calculation remains limited to an idealized seven-segment monochromatic model and should therefore be interpreted as a demonstration of application specific reference design rather than a complete segment-phasing system.

\section{Discussion}
\label{sec:discussion}

The results presented here support a broader interpretation of the phantom ZWFS than the specific four-reference, core preserving realization used for the initial characterization. The defining element of the architecture is the introduction of deliberately generated off-axis coherent references whose positions, strengths, and phase shifts can be selected independently of the treatment of the central PSF core. As described in Section~\ref{sec:theory_reconstruction}, these references may replace the central ZWFS reference when the core is required for another optical function, or they may augment a conventional central phase-contrast measurement. The reference geometry therefore becomes an optical design variable rather than a fixed property of the sensor.

The four-reference realization considered in Sections~\ref{sec:four_reference_characterization} and~\ref{sec:benchmark_classical_vs_phantom} provides the simplest illustration of this design freedom. Two orthogonal pupil carriers generate four satellite references and produce a strongly structured Fourier response. The resulting angular anisotropy is a property of this particular geometry rather than of the phantom framework in general. In favourable orientations the sensing photons can be used with an efficiency approaching the classical-ZWFS reference under the adopted normalization, whereas oblique Fourier modes can be substantially less efficiently measured. The corresponding Zernike benchmark shows the same effect in a modal basis: some modes exhibit a broader or more nearly linear response, while others suffer increased cross-talk or remain more favourably reconstructed by the classical ZWFS. The four-reference configuration therefore redistributes wavefront information rather than providing a uniform improvement over a conventional sensor.

The segment-piston example provides a more application specific use of the same principle. Here the relevant wavefront subspace and the symmetry of the pupil are known in advance, so angularly uniform sensitivity is not the design objective. Instead, three carrier axes separated by $60^\circ$ generate six auxiliary references matched to the segmented aperture, while the conventional central ZWFS phase region is retained. The resulting hybrid therefore uses the engineered channels to supplement, rather than replace, the high-sensitivity central phase-contrast measurement.

Within the monochromatic seven-segment simulation, this additional optical diversity becomes increasingly useful away from the phased state. Most importantly, the improvement is also present in the detector level information content and is not solely a consequence of the nonlinear estimator. At $1.5$ rad RMS simultaneous piston, using $10^5$ total detected WFS photons per measurement for each sensor, the median local
conditional Poisson Cram\'er--Rao lower bound is approximately $0.0172$ rad RMS for the hybrid sensor and $0.0336$ rad RMS for the vector ZWFS. The corresponding median condition numbers of the Fisher information matrix are approximately $8.1$ and $14.1$. Because these quantities are calculated directly from the detector intensities and their derivatives with respect to the segment pistons, before carrier demodulation or nonlinear reconstruction, they show that the auxiliary references change the optical encoding itself and provide a better-conditioned local measurement at large piston excursions.

This result also clarifies how the anisotropy observed for the four-reference sensor should be interpreted. For a general purpose wavefront sensor, one may seek a sufficiently diverse reference geometry to distribute information approximately uniformly over the controlled spatial frequency range. For an application with a known error subspace, however, uniformity need not be optimal. The number, orientation, amplitude, and phase of the reference channels can instead be selected to improve the observability of the modes that matter for the instrument. The segmented-pupil case considered here is one example of such application specific reference design; other controlled modal subspaces could favour different geometries.

The core preserving photonic realization remains a complementary motivation. In an instrument where the diffraction-limited central PSF is itself the science input to a single mode fibre, waveguide, or photonic beam combiner, the off-axis references provide a means of performing phase-contrast sensing without requiring the physical on-axis core to serve as the sensing reference. This does not remove the science-versus-sensing photon trade. Generating the engineered references transfers a finite fraction of the incident power away from the central diffraction order, and the useful operating point must therefore balance science coupling, reference-channel flux, modal sensitivity, and reconstruction performance. The generalized formulation of Section~\ref{sec:theory_reconstruction} makes clear that this photon-sharing implementation is one possible realization rather than a requirement of the phantom architecture.

Several limitations remain before either realization can be assessed at instrument level. The present analysis is monochromatic and scalar, the focal plane phase masks and pupil modulation are assumed to be known exactly, and the photonic interface is represented by an overlap integral model rather than a full electromagnetic scattering calculation. The engineered satellite references are diffractive and will therefore move and broaden with wavelength, modifying both the focal plane phase interaction and the spatial demodulation in broadband operation. The reconstruction calculations also assume accurate calibration and do not yet include amplitude aberrations, optical mis-registration, detector systematics, or time dependent reference errors.

The segmented-piston example has additional application specific limitations. It considers an idealized seven-segment pupil, an equal detected-photon comparison, and pistons restricted to the principal monochromatic phase branch. The additional references improve fine and intermediate phasing within this branch but do not remove the fundamental $2\pi$ ambiguity; coarse acquisition would still require wavelength diversity or another absolute-phasing observable. The scalar and vector ZWFS benchmarks are likewise idealized optical models. The present result should therefore be interpreted as evidence that engineered references can improve the information geometry of a selected wavefront subspace, rather than as a complete segmented telescope phasing architecture.

The most direct next step is experimental validation of the underlying optical principle. A monochromatic bench experiment could generate selectable pupil carriers, apply localized focal plane phase shifts to the resulting satellite references, and compare the measured direct and demodulated pupil responses with the predicted forward model. Beyond this, broadband propagation and application specific optimization should treat the reference geometry, photon allocation, central phase-contrast element, and estimator jointly. Such studies will determine whether engineered reference phase contrast provides a practical advantage in photonic coupling, segmented-aperture phasing, or other instruments with strongly structured wavefront sensing requirements.

\section{Conclusion}
\label{sec:conclusion}

We have introduced the Phantom Zernike Wavefront Sensor as an engineered reference extension of phase-contrast wavefront sensing. Controlled pupil plane modulation generates off-axis coherent speckle structures that undergo localized focal plane phase shifts and modify the downstream pupil intensity. These engineered references may be used as the primary phase-contrast channels, leaving the physical on-axis PSF core available for another optical function, or may be added to a conventional central ZWFS response to provide auxiliary wavefront information.

Using a monochromatic scalar forward model, we derived the carrier structure of the resulting pupil plane signal and showed how the engineered references can be recovered through spatial demodulation and incorporated into a calibrated reconstruction. A simple four-reference realization demonstrates that reference geometry redistributes sensitivity across spatial frequency magnitude and orientation. Its strong fourfold angular dependence and mode dependent Zernike reconstruction behaviour show that this geometry does not provide a uniform improvement over a classical ZWFS; rather, it establishes that the optical response can be altered through the placement and phase of the coherent reference channels.

We then considered differential piston on a seven-segment aperture as a concrete application in which the relevant wavefront subspace and pupil geometry are known a priori. A hybrid configuration retaining the conventional central ZWFS phase region and adding six engineered references on three carrier axes preserves the high-sensitivity response near the phased state while providing additional optical diversity at larger piston excursions. At $1.5$ rad RMS simultaneous piston and $10^5$ total detected WFS photons per measurement, the median local conditional Poisson Cram\'er--Rao lower bound is approximately $0.0172$ rad RMS for the hybrid architecture compared
with $0.0336$ rad RMS for the vector ZWFS, with a corresponding improvement in the conditioning of the six-parameter Fisher information matrix. This detector level result demonstrates that the engineered references can modify the available wavefront information itself rather than merely providing an alternative reconstruction of the same optical signal.

The principal conclusion is therefore not that a particular phantom geometry replaces the classical ZWFS, but that deliberately generated coherent reference channels provide additional optical degrees of freedom with which a phase-contrast sensor can be matched to its application. The core preserving four-reference implementation and the hybrid segmented-pupil implementation represent two different uses of this framework: one separates the sensing reference from a central science channel, while the other augments an existing ZWFS response in a selected wavefront subspace.

Future work should test these principles experimentally and extend the model to broadband propagation, realistic optical interfaces, amplitude aberrations, calibration errors, and closed loop operation. More generally, optimization of an engineered reference sensor should be performed jointly over the reference geometry, phase shifts, photon allocation, central focal plane interaction, and the wavefront subspace relevant to the instrument. The phantom ZWFS therefore provides a framework for moving from a fixed phase-contrast reference geometry toward application specific wavefront sensing architectures.

\begin{acknowledgments}
B. Courtney-Barrer acknowledges support from JSPS KAKENHI grant No.22K21349 (International Leading Research by S. Miyazaki et al.).
\end{acknowledgments}

\bibliography{PASPsample701}{}
\bibliographystyle{aasjournalv7}

\appendix

\section{Carrier expansion, demodulation conditions, and reduced-envelope validity}

\subsection{First-order waffle expansion and retained carrier coefficients}
\label{sec:appendix_waffle}
This appendix makes explicit the retained first order carrier expansion used in Section~\ref{sec:theory_reconstruction}. Writing
\begin{equation}
\phi_{\mathrm{w}}(\mathbf{r}) = a\left[\sin(2\pi f x)+\cos(2\pi f y)\right],
\end{equation}
we factor the waffle term as
\begin{equation}
e^{i\phi_{\mathrm{w}}(\mathbf{r})}
=
e^{ia\sin(2\pi f x)} e^{ia\cos(2\pi f y)}.
\end{equation}
Using the Jacobi-Anger identities,
\begin{equation}
e^{ia\sin X}=\sum_{n=-\infty}^{\infty} J_n(a)e^{inX},
\qquad
e^{ia\cos Y}=\sum_{m=-\infty}^{\infty} i^m J_m(a)e^{imY},
\end{equation}
we obtain
\begin{equation}
e^{i\phi_{\mathrm{w}}(\mathbf{r})}
=
\sum_{n,m} J_n(a)\,i^m J_m(a)\,e^{i2\pi(nfx+mfy)}.
\end{equation}
For $a\ll 1$, the Bessel functions satisfy $J_0(a)=1+\mathcal O(a^2)$, $J_{\pm1}(a)=\pm a/2+\mathcal O(a^3)$, and $J_{|n|\ge2}(a)=\mathcal O(a^2)$. Retaining only terms through first order in $a$ gives
\begin{equation}
e^{i\phi_{\mathrm{w}}(\mathbf{r})}
\approx
1
+\frac{a}{2}e^{i2\pi f x}
-\frac{a}{2}e^{-i2\pi f x}
+\frac{ia}{2}e^{i2\pi f y}
+\frac{ia}{2}e^{-i2\pi f y}.
\end{equation}
The retained carrier coefficients are therefore
\begin{equation}
\beta_{0}=1,
\qquad
\beta_{x+}=\frac a2,
\qquad
\beta_{x-}=-\frac a2,
\qquad
\beta_{y+}=\frac{ia}{2},
\qquad
\beta_{y-}=\frac{ia}{2}.
\end{equation}
All second-order harmonics, including $(\pm 2f,0)$, $(0,\pm 2f)$, and oblique terms such as $(\pm f,\pm f)$, are neglected in the retained carrier model. These terms reappear in the full intensity expansion as higher frequency contributions that are rejected by the demodulation step when the carrier bands are adequately separated.

\subsection{Reduced envelope model and its regime of validity}
\label{app:reduced_envelope}

The notation introduced in Eqs. \ref{eq:Bq_def} and \ref{eq:Dq_def} compresses the exact optical propagation into a reduced carrier envelope model. This appendix states more explicitly what is contained in the operator \(L\), why it is linear in the optical field, and what assumptions are required for the carrier envelope approximation to be useful. Let $\widetilde E(\mathbf u)=\mathcal F\{E(\mathbf r)\}$ denote the
focal plane field produced by an arbitrary input pupil field $E(\mathbf r)$.
Consistent with the general formulation of Section~2.1, the complete linear
field-propagation operator is
\begin{equation}
\mathcal L[E](\mathbf r)
=
\mathcal F^{-1}
\left\{
M_0(\mathbf u)\,
M_s(\mathbf u)\,
\mathcal R
\left[
\mathcal F\{E\}
\right](\mathbf u)
\right\}.
\label{eq:app_general_L}
\end{equation}
Here $M_0$ denotes the optional central phase-contrast region, $M_s$ the
engineered satellite phase shifts, and $\mathcal R$ any fixed linear routing
or filtering operation acting in the focal plane. Each of these operations is
linear in the complex optical field: $\mathcal R$ is assumed linear by
construction, while $M_0$ and $M_s$ act by multiplication by fixed complex
functions. Fourier propagation is likewise linear. The operator
$\mathcal L$ is therefore linear in the optical field even though the
subsequent detector measurement,
\begin{equation}
I_{\rm out}(\mathbf r)
=
\left|
\mathcal L[E](\mathbf r)
\right|^2,
\end{equation}
is nonlinear.

The core preserving four-reference realization used in Sections~3 and~4 is a
specific case of this general operator. For that realization the central
phase region is absent, $M_0=1$, and the routing operator is the photonic
operator $\mathcal R_{\rm chip}$ introduced in Section~2.2. For an arbitrary
focal plane field $\widetilde E$, this is
\begin{equation}
\mathcal R_{\rm chip}
\left[
\widetilde E
\right](\mathbf u)
=
r
\left[
(1-c(\mathbf u))\widetilde E(\mathbf u)
+
c(\mathbf u)
\left(
\widetilde E(\mathbf u)
-
\left\langle
\widetilde E,g
\right\rangle
g(\mathbf u)
\right)
\right],
\label{eq:app_Rchip}
\end{equation}
where
\begin{equation}
\left\langle
\widetilde E,g
\right\rangle
=
\int
\widetilde E(\mathbf u)g^*(\mathbf u)\,d\mathbf u.
\label{eq:app_overlap}
\end{equation}
The corresponding field-propagation operator is therefore
\begin{equation}
\mathcal L_{\rm cp}[E]
=
\mathcal F^{-1}
\left\{
M_s\,
\mathcal R_{\rm chip}
\left[
\mathcal F\{E\}
\right]
\right\},
\label{eq:app_Lcp}
\end{equation}
which is the specialization of Eq.~(\ref{eq:app_general_L}) used for the
standalone core preserving benchmark. The subtraction of the projection onto
the guided mode, multiplication by the fixed Fresnel coefficient, focal plane
masking, and Fourier propagation are all linear operations on the complex
field. The further small signal approximation introduced in Section~2.4 is
instead the linearization of the nonlinear detector intensity about the
zero aberration reference field.

The reduced carrier envelope approximation used in Eqs.~\ref{eq:Bq_def} and \ref{eq:Dq_def} is a separate approximation. In the weak waffle description, the incident field is decomposed into a central component and a small number of shifted carrier components,
\begin{equation}
    E(r) \simeq \sum_{q\in Q} \beta_q P(r)e^{i2\pi k_q\cdot r}.
\end{equation}
The approximation assumes that, after propagation through the focal plane coupling, local speckle phase shifts, and return to the pupil, the contribution from each retained carrier can still be represented as that same rapidly varying carrier multiplied by a slowly varying complex envelope:
\begin{equation}
    L\left[\beta_qP e^{i2\pi k_q\cdot r}\right]
    \approx
    B_q(r)e^{i2\pi k_q\cdot r},
\end{equation}
and, for the aberration dependent perturbation,
\begin{equation}
    L\left[i\beta_qP\phi\, e^{i2\pi k_q\cdot r}\right]
    \approx
    D_q(r)e^{i2\pi k_q\cdot r}.
\end{equation}
The envelopes \(B_q\) and \(D_q\) absorb the effects of the finite pupil, the focal plane phase mask shape, partial coupling or reflection at the waveguide interface, and any slowly varying amplitude and phase structure introduced by the return optics. They are not assumed to be known analytically. Their role is to make explicit the carrier structure that motivates the demodulation.

This approximation is expected to be most accurate when the following conditions hold. 
First, the waffle amplitude should be small enough that the central and first order carrier terms dominate the field. Higher order waffle harmonics are then weak compared with the retained orders. If the waffle amplitude is increased, higher order Bessel terms produce additional satellite orders and additional beat terms in the returned pupil intensity. These terms are included in the exact numerical propagation used for the simulations, but they are not represented in the first order analytic carrier model.

Second, the phase shifts applied to each speckle should be spatially localized around the first order satellite PSF copies and should remain separated from the central PSF core and from one another. In this case each tag primarily modifies the intended satellite reference channel. If a tag overlaps the central core, it begins to behave partly like a conventional Zernike phase mask. If neighbouring satellite tags overlap, or if a tag is so large that it includes substantial flux from other diffraction orders, the simple association between one tag and one carrier channel breaks down.

Third, the carrier separation \(f\) must be large enough for the desired first order carrier terms to be isolated by spatial demodulation. After multiplication by \(e^{-i2\pi f x}\), the desired \(+x\) carrier is shifted to baseband, while the zero order, opposite carrier, second order terms, and oblique cross terms are shifted away from baseband. The low pass filter must therefore be chosen to pass the bandwidth of the slowly varying envelope while rejecting these displaced terms. If the carrier separation is too small, or if the envelopes are too broad in spatial frequency, the demodulated channels are contaminated by neighbouring carrier terms. 

Fourth, for the core preserving photonic realization specifically, the focal plane coupling region associated with the science core should not introduce strong mixing between the central core and the satellite reference channels.

Fifth, the focal plane phase shifts should generate a non degenerate complex reference response. The chosen phase progression is intended to produce both in phase and quadrature components in the demodulated pupil signal. Other phase shifting choices are possible, but if the returned carrier channels do not provide sufficient phase diversity, the interaction matrix becomes poorly conditioned and some modes become difficult to reconstruct.

The reduced envelope model should therefore be interpreted as an analytic description of the signal structure, not as a substitute for the full forward model. The numerical results in this paper are generated by propagating the full complex field through the scalar forward model defined in \ref{sec:forward_model}, including the full exponential waffle phase and the applied focal plane masks. The weak waffle carrier expansion and the envelope notation are used to explain why the returned intensity contains known spatial carriers and why mixing with those carriers followed by low pass filtering can recover useful linear observables.

\subsection{Conditions for valid carrier demodulation}
\label{sec:appendix_demodulation}
The demodulation model in Section~\ref{sec:theory_reconstruction} assumes that the retained first order carriers can be isolated by mixing and low-pass filtering. The main requirements are as follows.

First, the retained carrier bands at $(\pm f,0)$ and $(0,\pm f)$ must be sufficiently separated from the zero order term and from one another after propagation through the sensing branch. This requires the waffle frequency $f$ to be large enough that the baseband support of the slowly varying envelopes does not overlap adjacent translated copies after mixing.

Second, the low-pass filter must retain the envelope bandwidth while rejecting the translated zero order, second-order, and oblique terms. After multiplication by $e^{-i2\pi f x}$, for example, the desired $+x$ carrier is shifted to baseband, while the zero order term moves to $(\mp f,0)$, the $-x$ carrier to $(\mp 2f,0)$, and oblique terms to locations such as $(\mp f,\pm f)$. The filter cutoff must therefore be chosen to pass the support of $\Gamma_x(\mathbf r)$ while rejecting these displaced components. The same logic applies to the $y$ carrier.

Third, the envelopes $\Gamma_x$ and $\Gamma_y$ must vary slowly compared with the carrier oscillations. In practice this means that the downstream pupil plane modulation is well described as a carrier multiplied by a low-spatial frequency complex envelope rather than by a broad spectrum field. This is the sense in which the demodulation is a carrier envelope approximation.

Finally, the phase masks applied to the focal plane  satellite spots must generate a non-degenerate complex reference response. In practice, the relative phase progression across the tagged spots is chosen so that, after propagation back to the pupil plane, both in-phase and quadrature information are present in the demodulated observables. Without this phase diversity, one expects reduced sensitivity or increased degeneracy in the reconstructed mode coefficients.

\section{Classical ZWFS benchmark in parallel notation}
\label{sec:appendix_classical_parallel}
For comparison, the classical ZWFS benchmark used in Section~\ref{sec:benchmark_classical_vs_phantom} may be written in a parallel but simpler notation. Let $\mathcal L_{\mathrm Z}$ denote the linear optical operator corresponding to focal plane  phase shifting of the central PSF core and re-imaging to the pupil plane. The exact output field is
\begin{equation}
E_{\mathrm Z,out}(\mathbf r)=\mathcal L_{\mathrm Z}\!\left[P(\mathbf r)e^{i\phi(\mathbf r)}\right].
\end{equation}
For $|\phi|\ll 1$,
\begin{equation}
E_{\mathrm Z,out}(\mathbf r)
\approx
E_{\mathrm Z,out}^{(0)}(\mathbf r)+\delta E_{\mathrm Z,out}(\mathbf r),
\end{equation}
with
\begin{equation}
E_{\mathrm Z,out}^{(0)}(\mathbf r)=\mathcal L_{\mathrm Z}[P](\mathbf r),
\qquad
\delta E_{\mathrm Z,out}(\mathbf r)=\mathcal L_{\mathrm Z}[iP\phi](\mathbf r).
\end{equation}
The differential intensity is then
\begin{equation}
\Delta I_{\mathrm Z}(\mathbf r)
\approx
2\Re\!\left[\left(E_{\mathrm Z,out}^{(0)}\right)^*\delta E_{\mathrm Z,out}\right].
\end{equation}
Thus both the classical ZWFS and the phantom framework rely on the same
first-order interference principle: a known zero aberration reference field
interferes with a perturbation field linear in $\phi$. In the standalone
four-reference realization considered in Sections~3 and~4, the phase-contrast
reference is intentionally distributed among engineered off-axis carrier
channels and recovered by spatial demodulation. In the hybrid realization of
Section~5, these engineered channels instead augment the conventional central
ZWFS response.

\section{Full modal benchmark results}\label{sec:appendix_benchmark}
This appendix collects the full set of mode by mode benchmark results corresponding to the representative examples shown in Section \ref{sec:benchmark_classical_vs_phantom}. For each reconstructed Zernike mode, we show the recovered modal response as a function of injected amplitude, the corresponding residual phase RMSE, and the modal cross-talk structure inferred from the reconstructed coefficient vectors. These figures are intended to complement the main-text examples by showing explicitly that the relative performance of the phantom and classical ZWFS benchmarks is mode dependent: some modes benefit from the engineered reference geometry, while others remain more favourably reconstructed by the classical sensor.
\begin{figure}
    \centering
    \includegraphics[width=0.9\linewidth]{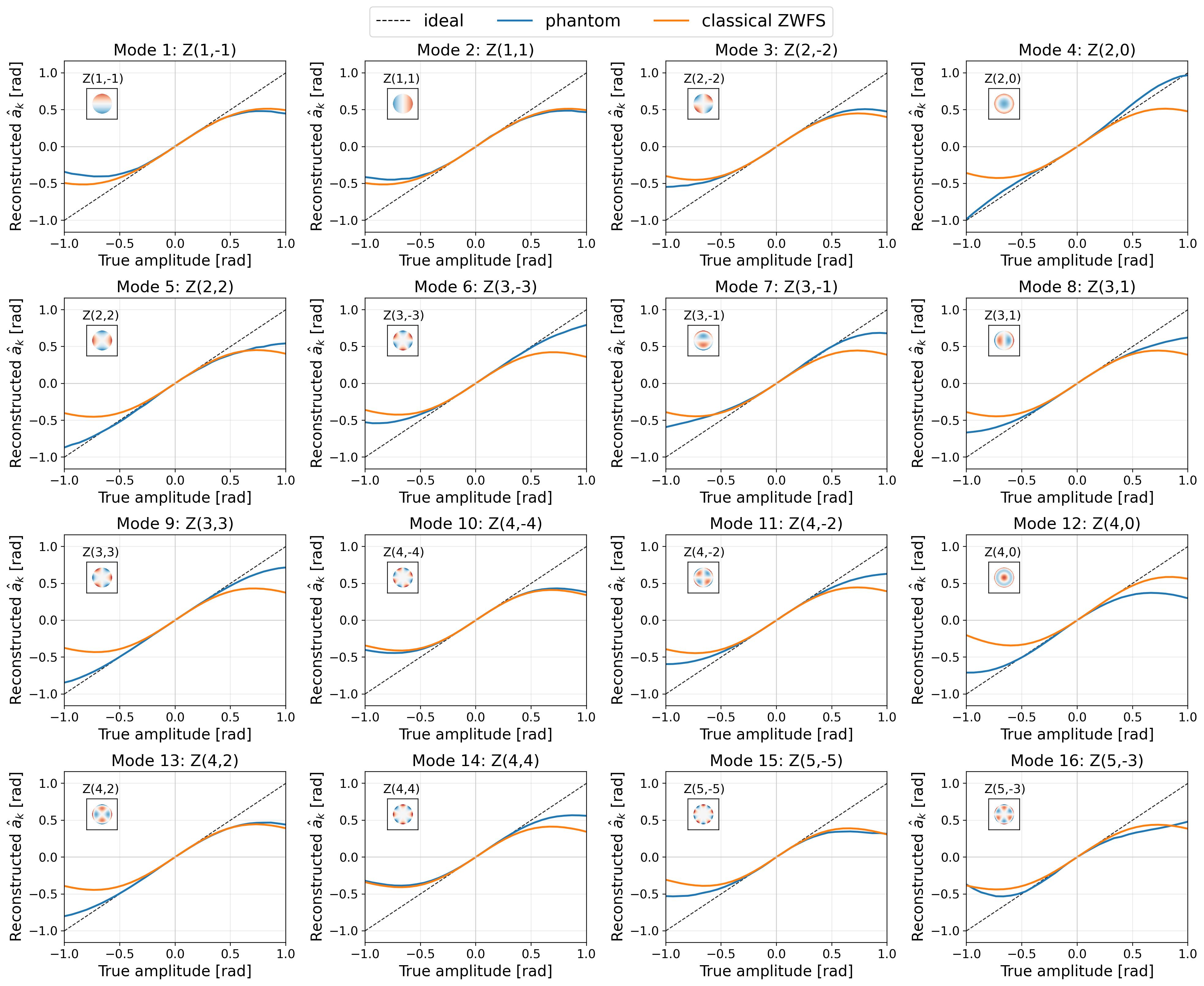}
    \caption{Full set of modal response curves for the phantom and classical ZWFS benchmarks.}
    \label{fig:appendix_all_mode_response}
\end{figure}

\begin{figure}
    \centering
    \includegraphics[width=0.9\linewidth]{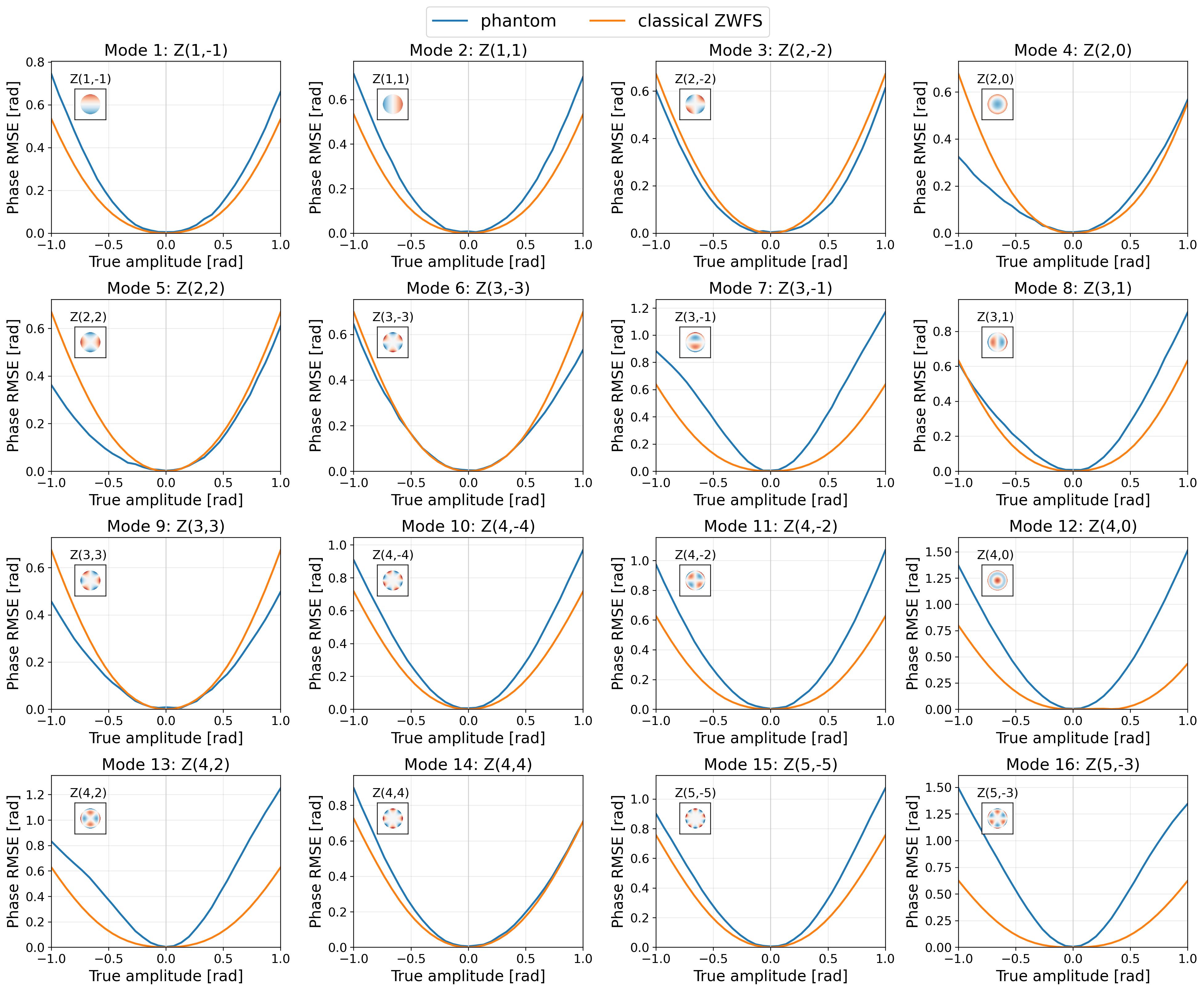}
    \caption{Full set of modal residual phase RMSE curves for the phantom and classical ZWFS benchmarks.}
    \label{fig:appendix_all_mode_rmse}
\end{figure}

\begin{figure}
    \centering
    \includegraphics[width=0.48\linewidth]{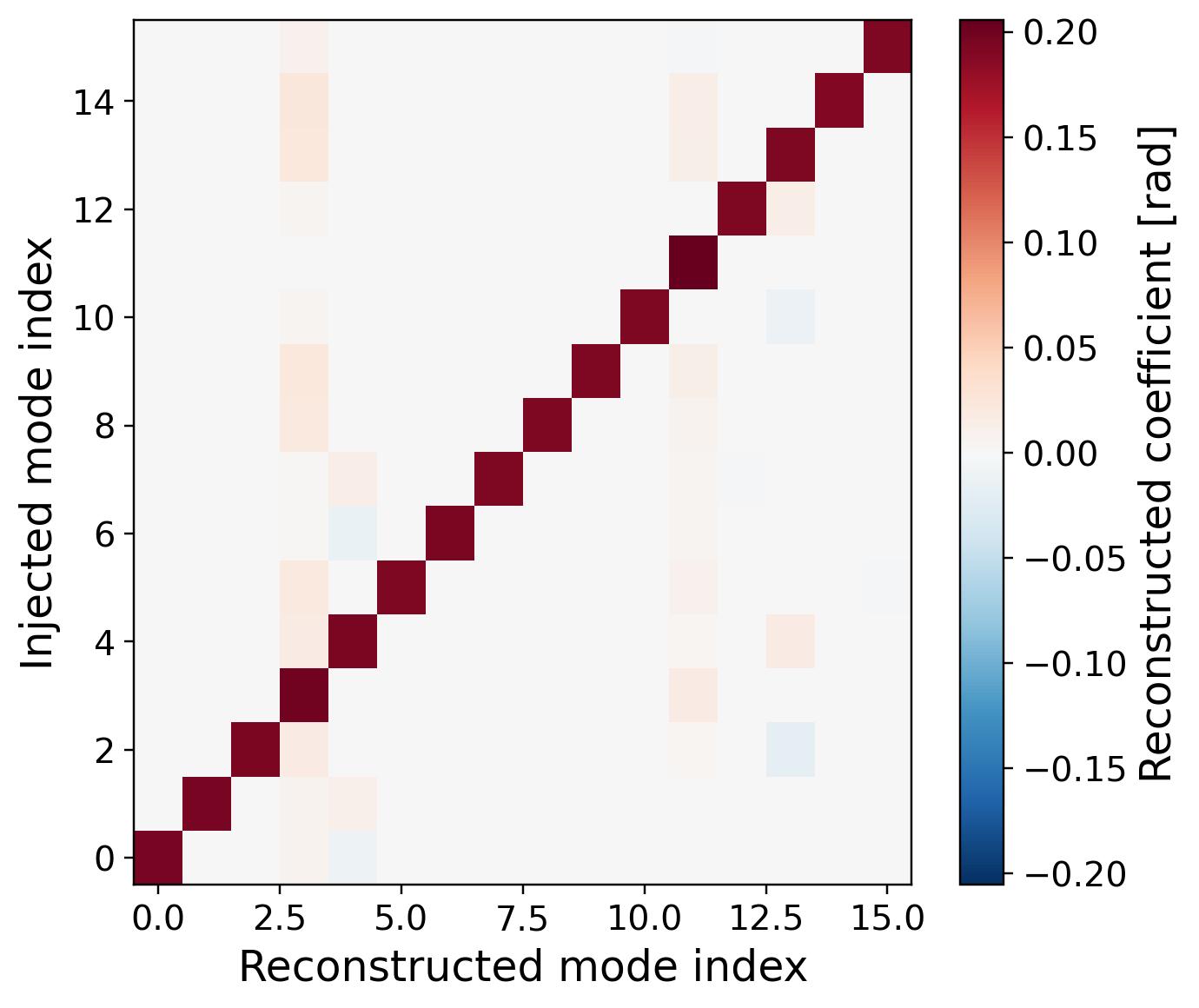}
    \includegraphics[width=0.48\linewidth]{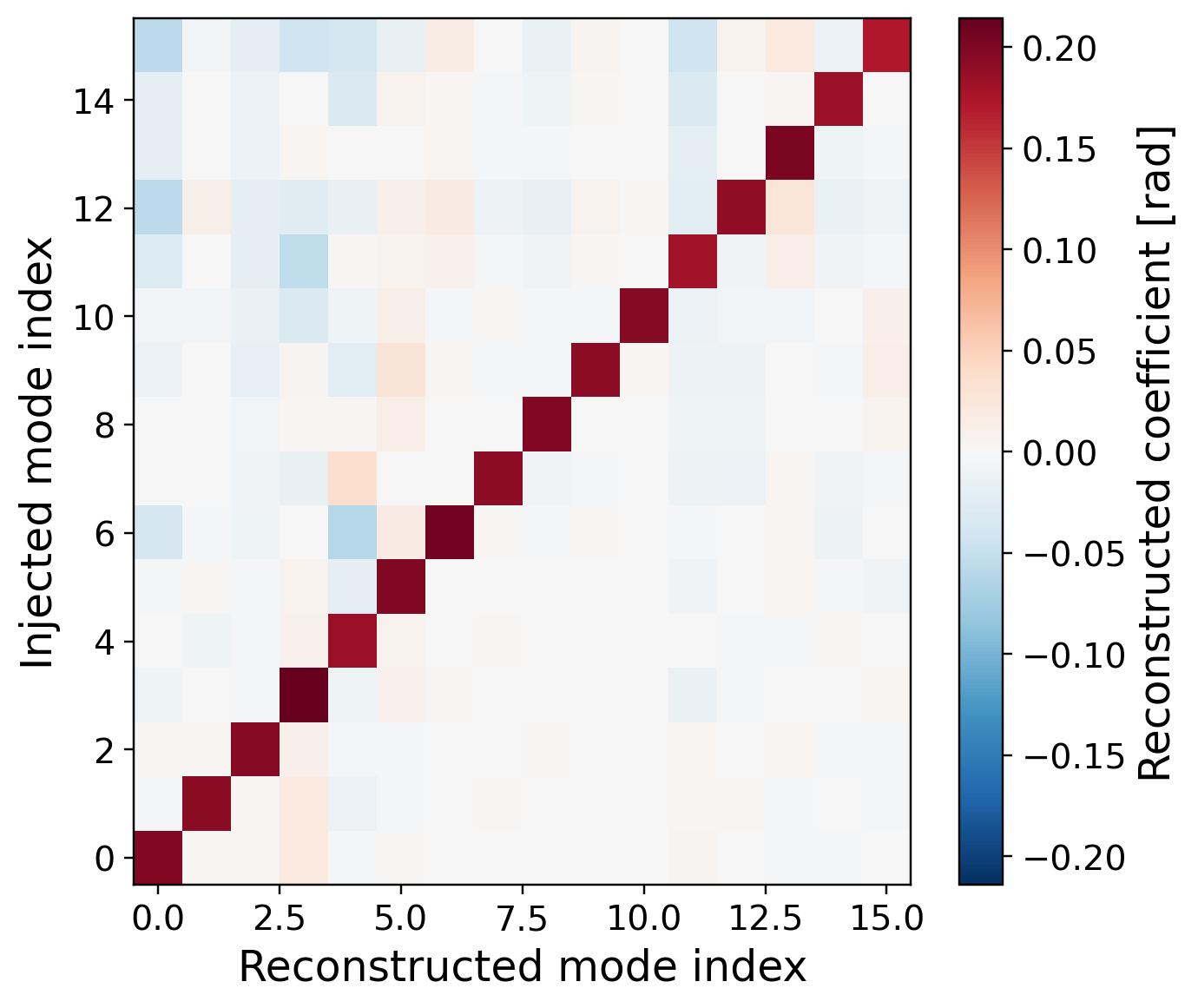}
    \caption{Modal cross-talk matrices at an injected amplitude of approximately \(0.2\) rad for the classical ZWFS (left) and phantom ZWFS (right).}
    \label{fig:appendix_crosstalk}
\end{figure}

\section{Detector level Fisher information for segment piston}
\label{app:segment_fisher}

The Fisher-information calculation used in
Section~\ref{sec:segment_piston_example} is performed directly on the expected
detector intensities and is independent of the carrier demodulation and
nonlinear piston estimator. The purpose is to compare the local information
content of the optical measurements under a common detected-photon
normalization.

Let $I_i(\mathbf p)$ denote the expected intensity in detector pixel $i$ for
piston vector $\mathbf p$, evaluated over a common pupil plane analysis
support. We define the normalized spatial photon distribution
\begin{equation}
    \pi_i(\mathbf p)
    =
    \frac{I_i(\mathbf p)}
    {\sum_{j\in{\rm sup}} I_j(\mathbf p)}.
    \label{eq:app_pi}
\end{equation}
The comparison adopts the same total detected WFS photon budget
$N_{\rm WFS}$ for each sensor. Because only the common analysis support is
used in the Fisher calculation, the expected number of photons contributing
to the measurement is
\begin{equation}
    N_{\rm sup}
    =
    N_{\rm WFS}
    \frac{
        \sum_{i\in{\rm sup}} I_i
    }{
        \sum_{i\in{\rm full}} I_i
    }.
    \label{eq:app_Nsup}
\end{equation}

Conditional on this detected photon number, the Fisher information matrix for
the six differential pistons is
\begin{equation}
    F_{st}
    =
    N_{\rm sup}
    \sum_{i\in{\rm sup}}
    \frac{1}{\pi_i}
    \frac{\partial\pi_i}{\partial p_s}
    \frac{\partial\pi_i}{\partial p_t}.
    \label{eq:app_fisher}
\end{equation}
This form measures the piston information contained in the spatial
distribution of the detected photons under the equal detected-WFS-photon
convention. It therefore does not include additional information that could
arise solely from changes in the total detected photon rate.

The derivatives are obtained analytically from the propagated complex field.
Using
\begin{equation}
    E_i(\mathbf p)
    =
    \sum_{s=0}^{6}
    E_{i,s}e^{ip_s},
\end{equation}
the field derivative with respect to piston $p_s$ is
\begin{equation}
    \frac{\partial E_i}{\partial p_s}
    =
    iE_{i,s}e^{ip_s},
\end{equation}
and hence
\begin{equation}
    \frac{\partial I_i}{\partial p_s}
    =
    2\Re\left[
        E_i^*
        \frac{\partial E_i}{\partial p_s}
    \right].
    \label{eq:app_dIdp}
\end{equation}
The derivative of $\pi_i$ then follows directly from
Eq.~(\ref{eq:app_pi}).

To avoid numerical singularities at exactly dark detector pixels, a uniform
background corresponding to a fractional intensity of $10^{-8}$ is included
when evaluating the Fisher matrix. This value is sufficiently small that it
acts only as a numerical regularization of zero-intensity pixels. The
resulting Fisher matrix is symmetrized before eigendecomposition.

The scalar precision reported in
Fig.~\ref{fig:segment_hybrid_performance}(b) is
\begin{equation}
    \sigma_{\rm CRLB}
    =
    \sqrt{
        \frac{1}{6}
        \operatorname{tr}
        \left(
            \mathbf F^{+}
        \right)
    },
\end{equation}
where $\mathbf F^{+}$ is the Moore--Penrose inverse. This quantity is the RMS
local Cram\'er--Rao lower bound averaged over the six differential piston
coordinates. The conditioning shown in
Fig.~\ref{fig:segment_hybrid_performance}(c) is
\begin{equation}
    \kappa_F
    =
    \frac{\lambda_{\max}(\mathbf F)}
         {\lambda_{\min}(\mathbf F)},
\end{equation}
with numerical eigenvalues below the adopted inversion tolerance excluded in
the same manner as for the pseudoinverse.

Both quantities are local properties of the measurement about a specified
piston state. They quantify precision and conditioning within a single
monochromatic phase branch, but do not describe the global capture range or
remove the intrinsic $2\pi$ ambiguity of monochromatic segment piston.

\end{document}